\documentclass[twocolumn]{aastex631}

\usepackage{amsmath}
\usepackage{subfigure}
\usepackage{xcolor}

\shorttitle{Planets in rings}
\shortauthors{Lee, Fuentes \& Hopkins}
\begin{document}

\title{Establishing Dust Rings and Forming Planets within Them}

\author[0000-0002-1228-9820]{Eve J. Lee}
\affiliation{Department of Physics and McGill Space Institute, McGill University, 3600 rue University, H3A 2T8 Montreal QC, Canada}
\affiliation{TAPIR, Mailcode 350-17, California Institute of Technology, Pasadena, CA 91125, US}

\author[0000-0003-2124-9764]{J. R. Fuentes}
\affiliation{Department of Physics and McGill Space Institute, McGill University, 3600 rue University, H3A 2T8 Montreal QC, Canada}

\author[0000-0003-3729-1684]{Philip F. Hopkins}
\affiliation{TAPIR, Mailcode 350-17, California Institute of Technology, Pasadena, CA 91125, US}

\correspondingauthor{Eve J. Lee}
\email{evelee@physics.mcgill.ca}

\begin{abstract}

Radio images of protoplanetary disks demonstrate that dust grains tend to organize themselves into rings. These rings may be a consequence of dust trapping within gas pressure maxima wherein the local high dust-to-gas ratio is expected to trigger the formation of planetesimals and eventually planets. We revisit the behavior of dust near gas pressure perturbations enforced by a planet in two-dimensional, shearing box simulations. While dust grains collect into generally long-lived rings, particles with small Stokes parameter $\tau_s < 0.1$ tend to advect out of the ring within a few drift timescales. Scaled to the properties of ALMA disks, we find that rings composed of larger particles ($\tau_s \geq 0.1$) can nucleate a dust clump massive enough to trigger pebble accretion which proceeds to ingest the entire dust ring well within $\sim$1 Myr. To ensure the survival of the dust rings, we favor a non-planetary origin and typical grain size $\tau_s \lesssim 0.05$--0.1. Planet-driven rings may still be possible but if so we would expect the orbital distance of the dust rings to be larger for older systems.

\end{abstract}

\section{Introduction}
\label{sec:introduction}

Planets are born in disks of gas and dust around a central star. Despite the vast progress in understanding the process of planet formation, the earliest phases remain unclear \citep[e.g.,][]{2018haex.bookE.135A}. In particular, the physical processes behind the growth of micrometer-sized dust particles into rocky bodies with sizes of thousands of kilometers remain unresolved. 

One of the major obstacles in the coagulation of large solid bodies is the rapidity at which dust grains drift onto the central star due to aerodynamic drag \citep{1972fpp..conf..211W,1977MNRAS.180...57W}. In typical circumstellar disks around Sun-like stars, one-meter grain at one astronomical unit will be dragged to the inner disk edge within $\sim $200 years, approximately 4$-$5 orders of magnitude shorter than the typical disk lifetime \citep[e.g.,][]{2010AREPS..38..493C}. While CO measurements reveal the gas disk to appear larger than the dust disk probed in radio continuum \citep[e.g.,][]{Ansdell18,Long22}, suggesting the grains undergo some degree of radial drift \citep[e.g.,][]{Birnstiel14}\footnote{We note however that the size discrepancy between the CO gas and the continuum dust emission may arise from different optical depths even in the absence of radial drift \citep{Facchini17, Trapman19}.}, the fact that these dust disks are extended to a few 10s of AU implies that the drift must be halted or delayed.

The classic calculation of radial drift derives from the assumption of a smooth gas disk. Substructures such as local pressure maxima within gas can act as traps collecting inflowing dust grains into ring-like structures \citep[e.g.][]{Pinilla12}. In fact, most of the bright protoplanetary disks imaged by the Atacama Large Millimeter/submillimeter Array (ALMA) shows concentric rings \citep{2015ApJ...808L...3A,Andrews18}. What produces these pressure maxima is an open question \citep[see e.g.,][for a review]{2017ASSL..445...91P}. Some suggestions include anticyclonic vortices in the gas \citep[e.g.,][]{2014MNRAS.437..575L}, edges of a gap carved out by perturbation from massive planets \citep[e.g.,][]{2012ApJ...755....6Z, Dong17}, ice lines where volatiles condense \citep[e.g.,][]{2008A&A...480..859B}, and magnetic zonal winds \citep[e.g.,][]{2013ApJ...763..117D,Suriano17,Hu22}. Although their origin is not well understood, since dust can collect within pressure traps, they have been proposed to be the regions where planetesimals and eventually planets form. 

But can dust near and within pressure maxima be collected into sufficiently high density clumps to trigger secondary instabilities (e.g., streaming instability; \citealt{Youdin05,Johansen07,2020MNRAS.498.1239S}) and/or collapse into bound planetesimals via self-gravity \citep{2010AREPS..38..493C,simon16,Gerbig20}? 
Using 1D (radial) and 2D (radial-vertical) hydrodynamic simulations, \citet{2016A&A...591A..86T} found that once the dust grains collect within a pressure bump and reach a local dust-to-gas ratio of $\sim$ 1, the dust backreaction destroys the pressure bump within $\sim$ 500 orbital periods, suggesting that any long-lived gas/dust substructure as a viable site of planetesimal and planet formation requires continuous forcing (see however \citealt{2017EP&S...69...50O} for a different view, who report that dust clumps of sufficiently high density can undergo gravitational instability but away from the disk midplane). With more sophisticated 3D hydrodynamic simulations of a gas pressure bump that is continuously reinforced including dust grains and dust self-gravity, \citet{2021AJ....161...96C} find that particles can clump to the Roche density (and therefore be expected to collapse into planetesimals) robustly and efficiently through the action of the streaming instability over multiple bump widths of their simulation box, although whether it is the streaming or the gravitational instability that ultimately creates planetesimals may depend on the size of the grains \citep{Carrera22}.\footnote{It is not surprising that the streaming instability is active away from the formal center of the pressure bump, since at the bump center, the dust-gas relative velocity would approach zero, likely deactivating any drag-induced instabilities.}

Once these planetesimals coagulate within a dust ring, would they grow into massive bodies quickly enough to spawn gas giants? \citet{Morbidelli20} provided analytic arguments applying the theory of pebble accretion \citep{Ormel10,Lambrechts12} to the B77 ring in the Elias 24 system in \citet{2018ApJ...869L..46D}. Under the assumption that the dust rings are created by a Gaussian pressure bump, \citet{Morbidelli20} concluded that a 0.1$M_\oplus$ seed core can only grow up to $\lesssim$1$M_\oplus$ within these dust rings, mostly because of the large orbital distances where the dynamical timescales are long. The final core mass is even smaller if the seed is situated sufficiently far away from the center of the dust ring, where the dust density would be significantly lower.

In this work, we revisit the question of planet formation in dust rings. Our approach differs from and extends previous work in several important ways. First, while we focus primarily on 2D local, shearing box simulations, we investigate the dust-gas interaction in the $R-\phi$ (radial-azimuthal) plane rather than in the $R-z$ (radial-vertical) plane of the disk (i.e.\ we do not assume axisymmetry), under the assumption that the gravitational settling to the midplane occurs over much shorter timescale than any dynamical timescale on the plane of the disk (verified with a small number of explicit 3D simulations). Second, instead of initializing our simulation boxes with a pre-determined amount of dust particles distributed uniformly throughout the disk, we supply them over time from one side of the box, to simulate the drift of dust from the outer disk into a site of gas pressure bump, allowing (in principle) for an arbitrarily large buildup of dust mass as required in many models. Third, instead of simply imposing a pressure bump as an initial condition (where it would represent a purely transient effect and may not be able to act efficiently), we model it as an explicit acceleration term acting on the gas by a gravitational force, which mimics the presence of an embedded planet in the disk. And fourth, we extensively consider the subsequent evolution of dense dust rings and bound clumps in simulated bumps, including comparison to observations of dust rings, and the ability of clumps to collapse under self-gravity (including shear and diffusion/turbulence effects).

This paper is organized as follows. Section \ref{sec:methods} describes the model and numerical simulations used in this work. Further, we discuss the conditions for trapping particles in terms of the shape of the pressure bump and dust properties. In Section \ref{sec:planet_results}, we investigate the dust distributions and trap efficiencies in simulations with bumps due to a perturbation by a planet. In Section \ref{sec:rings}, we use the trap efficiencies from the simulations to estimate the expected mass evolution of the axisymmetric ring as well as the masses of the densest bound clumps within the rings we simulate (and compare to observational constraints). In Section \ref{sec:clump_evol_acc}, we investigate the ability of clumps to form and collapse under self-gravity, and the expected mass growth of the densest clumps after said collapse via pebble accretion. Finally, we summarize and conclude in Section \ref{sec:conclusions}.

\section{Problem \&\ Methods} \label{sec:methods}

\subsection{Problem Setup \&\ Equations Solved}
\label{ssec:equations}

We investigate the dynamics of dust grains near and at local pressure perturbations in a gas disk, established by a tidal interaction with a planet. 
To concentrate on the local dynamics, we adopt the ``shearing-box'' approximation, i.e., calculations are performed on a small Cartesian patch of the disk defined in a rotating frame centered on $(R_{0},\,\phi_{0} + \Omega_0\,t,0)$, where $\Omega_{0}=\Omega(R_{0})$ is the Keplerian orbital frequency at $R_{0}$. In this frame, the locally-Cartesian coordinates are ${\bf x} = (x,\,y,\,z)=(R-R_{0},\,R_{0}(\phi - \phi_0) - R_{0}\Omega_{0}\,t,\,z)$. Expanding the equations of motion to $\mathcal{O}(|R-R_{0}|/R_{0} \ll 1)$ gives the momentum equation for gas:
\begin{align}
\nonumber \frac{D{\bf u}}{Dt} = -\frac{\nabla P}{\rho} - 2\Omega_{0}\,\hat{\bf z} \times {\bf u} + 3\Omega_{0}^{2}\,{\bf x} - \Omega_{0}^{2}\,{\bf z} + {\bf a}_{\rm dust} + {\bf a}_{\rm bump}\, ,
\end{align}
where $D/Dt = \partial/\partial t + ({\bf u}\cdot\nabla)$ is the Lagrangian derivative, $\rho$ is gas density, $P$ is gas pressure, ${\bf u}$ is gas velocity, ${\bf a}_{\rm dust}$ is the ``back-reaction'' acceleration from the force of gas drag on grains (defined below), and ${\bf a}_{\rm bump}$ is an acceleration due to an imposed force that models the pressure bump. For simplicity, in all our calculations we consider an inviscid gas described by an isothermal equation of state $P=\rho\,c_{s}^{2}$ with $c_s$ the constant gas sound speed. 

The simulation box is initially uniform in density (and therefore uniform in pressure) and the initial velocity field is set to the equilibrium solution in the absence of dust ``back-reaction'' (${\bf a}_{\rm dust} \rightarrow \mathbf{0}$), and in the absence of a planet:
\begin{align}
\label{eqn:P.eqm} \bar{P} &= P_{0}, \\ 
 \nonumber \bar{\bf u} &= \left( 0,\ -\frac{3}{2}\,x\,\Omega_{0} - \eta\,U_{K},\ 0 \right)\, , \\ &=  \left( 0,\  -\frac{3}{2}\,x\,\Omega_{0} - \Pi\,c_{s},\ 0 \right)\, , \label{eqn:u.eqm}
\end{align}
where $P_{0}=P_{0}(R_{0})$ is the unperturbed gas pressure evaluated at the center of the simulation box,
$U_{K} \equiv \Omega_{0}\,R_{0}$ is the Keplerian velocity at the center of the simulation box, $\eta \equiv \eta(R=R_{0}) = -(\partial P_{0}(R)/\partial R) / (2\,\rho_{0}\,\Omega^{2}\,R) \equiv \tilde{\eta}\,(c_{s}/U_{K})^{2}$ is the usual dimensionless pressure support parameter (defined at $R_{0}$ for the disk profile {\em without} a bump), and $\Pi \equiv \eta\,(U_{K}/c_{s}) = \tilde{\eta}\,(c_{s}/U_{K})$.

Next, we consider the gravitational perturbation by a planet of mass $M_p$ located at ($x_p$, $y_p$, $z_p$), whose gravitational field only acts on the gas (we turn off planet's gravity on dust grains so that we can isolate the effect of dust-gas dynamics in the presence of perturbations in the underlying disk gas).\footnote{We verify with a limited set of simulations with planet's gravity on dust grains turned on that the overall qualitative behavior of the dust grains do not change.} In general, this planet will drive a wave \citep{1980ApJ...241..425G,1986ApJ...309..846L}, and planets that are massive enough will carve out a gap \citep[e.g.,][]{2002ApJ...572..566R,2012ARA&A..50..211K} in the vicinity of its orbit, creating a pressure bump located a few pressure scale heights away \citep{2017ApJ...835..146D}. We write the bump acceleration as
\begin{align}
{\bf a_{\rm bump}}= 2\,\Pi\,c_{s}\,\Omega_{0}{\bf \hat x} -\nabla \Phi_p\, ,
\end{align}
where the first term on the right hand side takes into account the acceleration due to the large-scale gas pressure gradient because the shear-periodic boundaries do not otherwise allow a pressure discontinuity between the $\pm \hat{\bf x}$ boundaries, and $\Phi_p$ is the planet's gravitational potential
\begin{align}
\Phi_p = - \frac{G M_p}{\sqrt{(x-x_p)^2 + (y-y_p)^2 + (z-z_p)^2 + r_s^2}}\, ,
\end{align}
where we introduce a smoothing length parameter, $r_s = 0.1$ (in the unit of the disk scale height), to avoid the divergence of the planet's gravitational attraction.

In the shearing-box approximation, the momentum equation for dust particles is
\begin{align}
\label{eqn:dust.eom} \frac{d {\bf v}}{d t} &= -\frac{{\bf v}-{\bf u}}{t_{s}} - 2\Omega_{0}\,\hat{\bf z} \times {\bf v} + 3\Omega_{0}^{2}\,{\bf x} - \Omega_{0}^{2}\,{\bf z}\, ,
\end{align}
where {\bf v} is the dust particle velocity and $d{\bf v}/dt$ is its Lagrangian derivative. We can thus write the acceleration on gas {\em from} dust grains (i.e.\ the ``back-reaction'' force on gas) as
\begin{align}
{\bf a}_{\rm dust} \equiv \frac{1}{\rho}\,\int \frac{d \rho_{\rm d}}{d^{3}{\bf v}}\,\left(\frac{{\bf v}-{\bf u}}{t_{s}} \right)\,d^{3}{\bf v} \,,
\end{align}
where $t_s$ is the stopping time of a {\em single} dust grain. We cannot assume that all grains at a given location move with the same velocity, so $d \rho_{\rm d}/d^{3}{\bf v}$ is the phase-space distribution of grains and $\rho_{d} \equiv \int (d \rho_{\rm d}/d^{3}{\bf v})\,d^{3}{\bf v}$ is the dust density. Since we are primarily interested in small grains, we assume an Epstein drag law such that the stopping time is given by
\begin{align} 
\label{eq:ts} t_{s} &\equiv \sqrt[]{\frac{\pi}{8}}\frac{\rho_{\rm grain} \,a_{\rm grain}}{\rho\,c_{s}}\, \left( 1+\frac{9\pi}{128} \frac{|{\bf v}-{\bf u}|^{2}}{c_{s}^{2}} \right)^{-1/2},
\end{align}
\citep{2006A&A...453.1129P} where $\rho_{\rm grain}$ and $a_{\rm grain}$ are the \textit{internal} grain density and radius, respectively. Because $t_{s}$ can depend on the local gas conditions (e.g.\ $\rho$), we define the usual dimensionless ``effective Stokes number'' $\tau_{s} \equiv \Omega_{0}\,t_{s}(\rho=\rho_{0},\,P=P_{0},\,R=R_{0},\,|{\bf v}-{\bf u}|=0) \approx 0.63 \,\rho_{\rm grain}\,a_{\rm grain}/\rho_{0}\,H$ where $\rho_{0} = P_{0}/c_{s}^{2}$, in terms of the value of $t_{s}$ evaluated for the equilibrium gas properties outside or absent the ``bump''. In steady-state {\em without} a bump and neglecting back-reaction on the gas, the dust equilibrium density is $\rho_{d} = \mu_{0}\,\rho_{0}$ (where $\mu_0$ is the equilibrium dust-to-gas mass ratio) with the Nakagawa-Sekiya-Hayashi drift velocities \citep{1986Icar...67..375N}:
\begin{equation}
\bar{\bf v} = \bar{\bf u} -(2\,\tau_{s},\,\tau_{s}^{2},\,0)\,\Pi\,c_{s}/(1+\tau_{s}^{2})\, .  \label{eq:vel_drift}
\end{equation}
While GIZMO has the capability to take the physical size and the internal density of the grains as input, we emphasize that our calculation is parametrized by the Stokes number $\tau_{s}$ and so the absolute values of $\rho$, $c_s$, $\rho_{\rm grain}$, or $a_{\rm grain}$ never directly factor into our simulations (in other words, our input parameter is $\tau_{s}$).

\subsection{Numerical Methods}

We integrate the equations described in Section \ref{ssec:equations} in {\small GIZMO} \citep{hopkins:gizmo},\footnote{A public version of the code, including all methods used in this paper, is available at \href{http://www.tapir.caltech.edu/~phopkins/Site/GIZMO.html}{\url{http://www.tapir.caltech.edu/~phopkins/Site/GIZMO.html}}} using the Lagrangian ``meshless finite mass'' (MFM) method for the hydrodynamics (validated in e.g.\ \citealt{hopkins:mhd.gizmo,hopkins:cg.mhd.gizmo,hopkins:gizmo.diffusion,su:2017.weak.mhd.cond.visc.turbdiff.fx}). Grains are integrated using the ``super-particle'' method \citep[see, e.g.][]{carballido:2008.grain.streaming.instab.sims,johansen:2009.particle.clumping.metallicity.dependence,bai:2010.grain.streaming.vs.diskparams,pan:2011.grain.clustering.midstokes.sims}, whereby the motion of each dust ``particle'' in the simulation follows equation \eqref{eqn:dust.eom}, but each represents an ensemble of dust grains with similar properties. Numerical methods for the integration are described and tested in \citet{hopkins.2016:dust.gas.molecular.cloud.dynamics.sims,lee:dynamics.charged.dust.gmcs,hopkins:2019.mhd.rdi.periodic.box.sims} with the back-reaction accounted for as in \citet{moseley:2019.acoustic.rdi.sims,seligman:2019.mhd.rdi.sims}, in a manner guaranteeing exact conservation. 

We initialize a box of side-length $L_{\rm box}$, with shear-periodic boundary conditions for gas \citep{1995ApJ...440..742H} and $N_{\rm 1D,\,gas}^{D}$ resolution elements, where $D$ is the number of dimensions. As described in Section \ref{ssec:equations}, the gas density is initially uniform within the box and the initial velocity field follows equation \ref{eqn:u.eqm}.
In all our calculations, we set $L_{\rm box}  = 6 H$ to capture the bump without degrading the physical resolution. To find the optimal resolution for our study, we increase gradually $N_{\rm 1D, \,gas}$ and find convergence in the results when $N_{\rm 1D,\, gas} = 128$. 

We set the mass of individual dust ``super-particles'' to be $m_{i,\,{\rm dust}} = 0.01\,\langle m_{i,\,{\rm gas}} \rangle = 0.01\,M_{\rm gas,\,box} / N_{\rm 1D,\,gas}^{D}$, where  $M_{\rm gas,\,box} $ is the total mass of the gas in the box. 
Dust grains enter the right side of the box $+\hat{\bf x}$ (i.e.\ $R > R_{0} + L_{\rm box}/2$) and exit the left side of the box $-\hat{\bf x}$ (i.e.\ $R < R_{0} - L_{\rm box}/2$). For the inflow boundary at $+\hat{\bf x}$, we spawn new dust particles on a $D-1$ dimensional mesh (with $N_{\rm 1D,\,gas}^{D-1}$ elements) at a constant rate, set to the equilibrium drift $\bar{\bf v}$, such that the steady-state dust flux into the box is ${\bf F} = \langle \mu_{0} \rangle\,\rho_{0}\,\bar{\bf v}$ with $\langle \mu_{0} \rangle = 0.01$ (so that, {\em without} a bump, the steady-state dust-to-gas ratio in the box is $\langle \mu_{0} \rangle$).\footnote{We stress that the dust flux is a ``nuisance parameter,'' as it only controls the rate of dust flowing into the bump, so changing it only changes the simulation time required for the bump to reach some interesting local dust-to-gas ratio.} We emphasize that the dynamically relevant quantity in our simulation is the dust-to-gas ratio---which enters into dust back-reaction---rather than the absolute mass of the gas (or dust) which is a wholly scalable quantity.

Since the time scale for vertical settling is short in comparison with the dynamical scales of interest in this work, we focus on 2D ($R-\phi$ or $x-y$) simulations. The 2D cases allow us to reach much higher resolution and are a plausible approximation for thin dust layers. %We will briefly discuss the dynamics we observe in a select few 3D simulations.
%, but potentially miss some dynamics.

\begin{figure*}
    \centering
    \includegraphics[width=\textwidth]{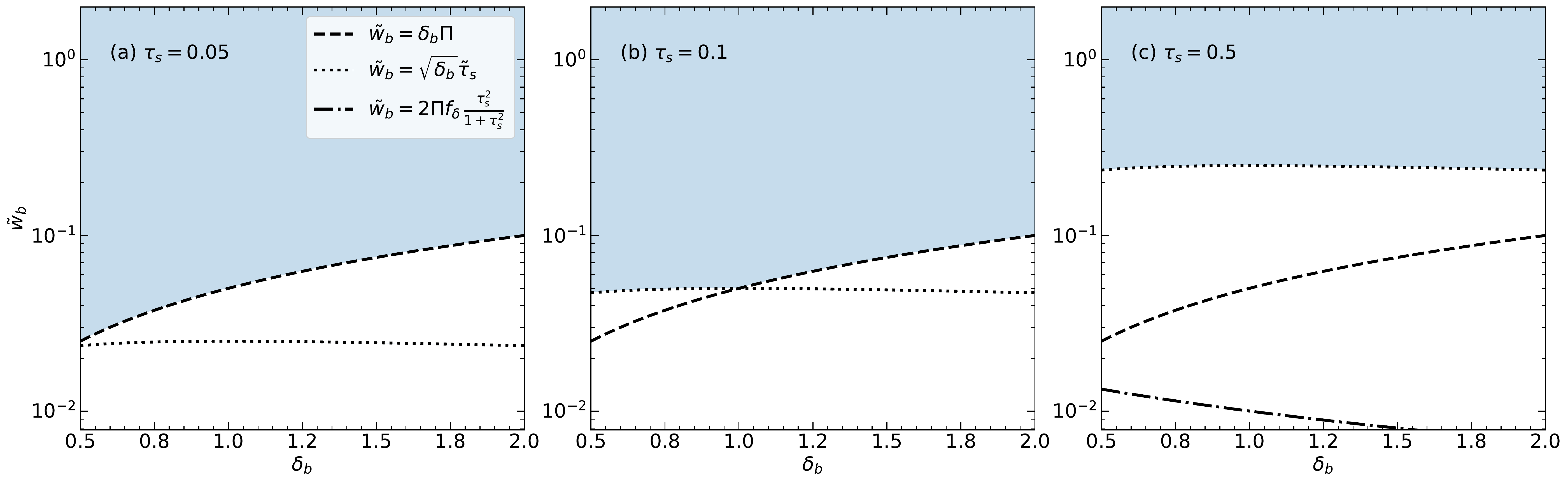}
    \caption{Parameter space constraints for the bump amplitude $\delta_b$ and width $\tilde{w}_b$, with $\Pi = 0.05$. The blue region represents the region of the parameter space that allows dust particles to be trapped by the bump, and the different lines denote the limits from the trapping requirements in Section \ref{ssec:param_space}. Since the size of the box is fixed to $L_{\rm box}/H = 6$, we limit the width of the bump up to a maximum of $\tilde{w}_b =2$, to ensure the bump is contained within the box. The line corresponding to the constraint $\tilde{w}_b = \delta_b/3.3\Pi$ is above $\tilde{w}_b = 2$ and therefore not shown here.}
    \label{fig:limits_tau}
\end{figure*}

\subsection{Parameter Space}
\label{ssec:param_space}

In our setup, there are three {\em physically meaningful} parameters: $\tau_{s}$, $\Pi$, and $M_p$. Other parameters either scale out entirely from the problem (e.g.\ absolute values of $\rho$, $c_{s}$, $R$, $\Omega$, etc.), or simply re-scale the rate of supply of dust, or are purely numerical parameters (e.g.\ dust and gas resolution, box size in units of $H$). 

Among the three parameters, $\Pi$ is narrowly constrained to $\sim$0.1: much larger values ($\Pi \gtrsim 1$) imply the ``disk'' is actually a quasi-spherical hydrostatic object, while much smaller ($\Pi \lesssim 0.01$) would automatically mean the disk has Toomre $Q<1$ {\em in the gas} and should fragment via gravitational instability. In all our simulations, $\Pi$ is set to 0.05. 

The planet mass $M_p$ controls the shape of the resulting pressure bump. For inviscid disks, \cite{1993prpl.conf..749L} showed that gas surrounding the planet can be marginally stable against Rayleigh's rotational instability when the Hill radius of the planet, $R_{\rm Hill} = r (M_p / 3M_{\star})^{1/3}$, is comparable to the disk scale height, $H$, which yields the ``thermal mass''
\begin{equation}
M_{\rm th} = \dfrac{2c_s^3}{3G\Omega_0}\, .
\end{equation}
We vary the mass of the planet in the interval $0.1 - 2.7M_{\rm th}$, and find that pressure bumps form when $M_p \gtrsim 0.5 M_{\rm th}$. As our simulation boxes are focused on a small local patch of the protoplanetary disk, we keep the mass of the planet $M_p < 2.5 M_{\rm th}$ to make sure the planet does not accrete significant amount of gas in the box. These constraints result in a narrow range of $M_p$, and we find that both the size of the bump and the dust dynamics do not change significantly so long as the pressure bumps are created. Therefore, we focus our discussion to $M_p = 2.25 M_{\rm th}$. We also fix the planet location at $x=-2H$ from the center of the simulation box so that the bump generated by the planet is contained in the box while giving enough room for dust particles to drift inward and interact with the bump.

Pressure bumps, once formed, need to be able to trap particles of a given $\tau_{s}$. By approximating the shape of the bump as a Gaussian
\begin{equation}
    P = P_{0} + P_{\rm bump} = P_{0}\,\left(1 + \delta_{b}\,e^{-x^{2}/2\,w_{b}^2} \right),
    \label{eq:Pbump}
\end{equation}
where $P_{\rm bump}$ is the resulting perturbation to the gas pressure because of planet's gravity,
we can identify the physically reasonable values of the bump amplitude $\delta_{b}$ and the bump width $w_{b}$ \textbf{to} determine whether our choice of $M_p$ produces a pressure bump that is strong enough for a range of $\tau_{s}$:

\begin{itemize}
\item[1)] To be effectively a bump, the acceleration \textbf{of} the Gaussian bump (i.e., $\rho^{-1}\partial P_{\rm bump}/\partial R$) has to be greater than \textbf{that of} the background pressure gradient (i.e., $\rho^{-1}\partial P_{0}/\partial R = 2\Pi c_s \Omega_{0}$), particularly at one sigma from the peak where the acceleration is greater than at larger distances. In other words, the pressure gradient of the bump at $x=w_{b}$ needs to be greater than the background gradient which results in
\begin{align}
\delta_{b} > 3.3\,\tilde{w}_b\,\Pi\,\, \text{ or}\, \,
\tilde{w}_b < \frac{\delta_{b}}{3.3\,\Pi} \, ,
\end{align}
where $\tilde{w}_b \equiv w_b/H$. Note that to guarantee an effective bump, we only need the exterior portion of the bump (i.e., where $a_{\rm bump} > 0$) to be steeper than the background gradient.

\item[2)] Dust particles need to slow down (on a time scale $\sim \tilde{t}_{s}$ ``stopping time in the bump'') before the equilibrium drift speed $v_{d,\,0}$ carries them ``through'' the bump (width $w_b$), i.e. $\tilde{t}_{s} \ll  t_{d,\,0} \sim w_b/v_{d,\,0}$. Using $v_{d,\,0} \sim 2\,\Pi\,c_{s}\,\tau_{s}/(1+\tau_{s}^{2})$, this requires
\begin{align}
\tilde{w}_b  \gtrsim 2\,\Pi\,\frac{\tau_{s}\,\tilde{\tau}_{s}}{1+\tau_{s}^{2}}  = 2\,\Pi\,f_{\delta}\,\frac{\tau_{s}^{2}}{1+\tau_{s}^{2}} 
\end{align}
where $\tilde{\tau}_{s}$ is the Stokes number ``in the bump'' with $\tilde{\tau}_{s} \approx  \tau_{s}/(1 + \delta_{b}) = f_{\delta}\,\tau_{s}$ in the Epstein regime.

\item[3)] The bump needs to ``catch'' grains accelerated by itself. As dust grains enter into the bump, assuming criteria above are met, grains accelerate up towards a new terminal velocity of $v_{d,\,b} \sim \tilde{t}_{s}\,\rho^{-1}\,\partial P_{\rm bump}/\partial R \sim 0.6\,\tilde{\tau}_{s}\,(\delta_{b}/\tilde{w})\,c_{s}$, crossing the peak (i.e.\ the ``trap region'' of width $w_b$) in a time $t_{d,\,b} \sim w_b/v_{d,\,b}$ which must be $<\tilde{t}_{s}$. Altogether, this gives
\begin{align}
\tilde{w}_b \gtrsim \delta_{b}^{1/2}\,\tilde{\tau}_{s}\, .
\end{align}
We note that condition 3 effectively describes the requirement to trap the particles once they cross over the peak into the inner side of the bump; it is also a more stringent condition than condition 2 in the limit $\tau_{s} < 1$ unless $\delta_b^{1/2} \leq 2\pi \tau_s$. In the parameter space we explore, if condition 3 is met, condition 2 is automatically met.

\item[4)] For the bump to be stable (i.e., the acceleration by the bump pressure gradient does not exceed Keplerian acceleration), we must have:
\begin{align}
\delta_{b} \lesssim 0.5\,\left( \frac{v_{K}}{c_{s}} \right)\,\tilde{w}_b \sim  \frac{\tilde{w}_b}{\Pi}.
\end{align}

\end{itemize}

\begin{figure*}
\centering
\includegraphics[width=8.1cm]{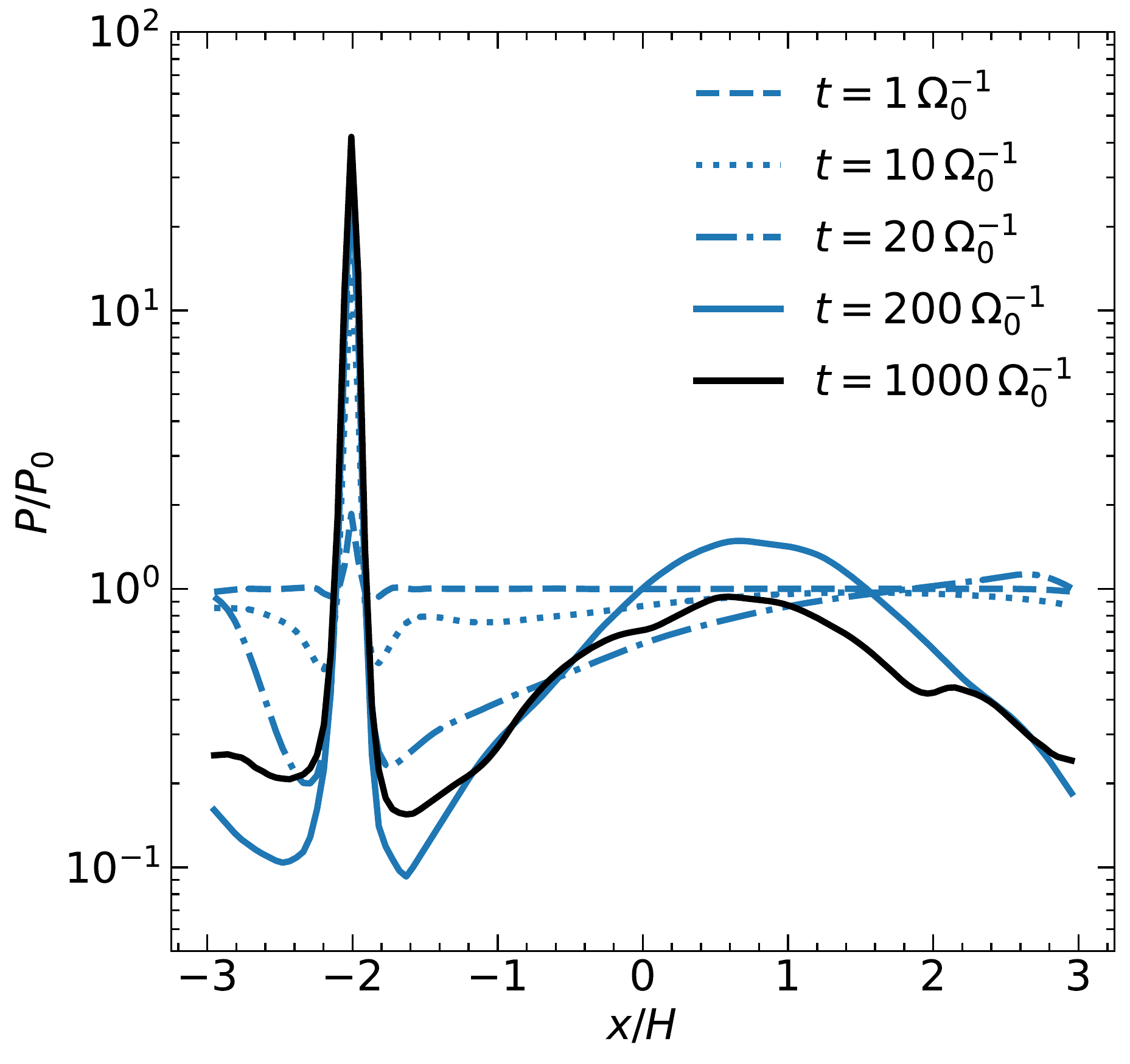} \hspace{1mm} \includegraphics[width=9cm]{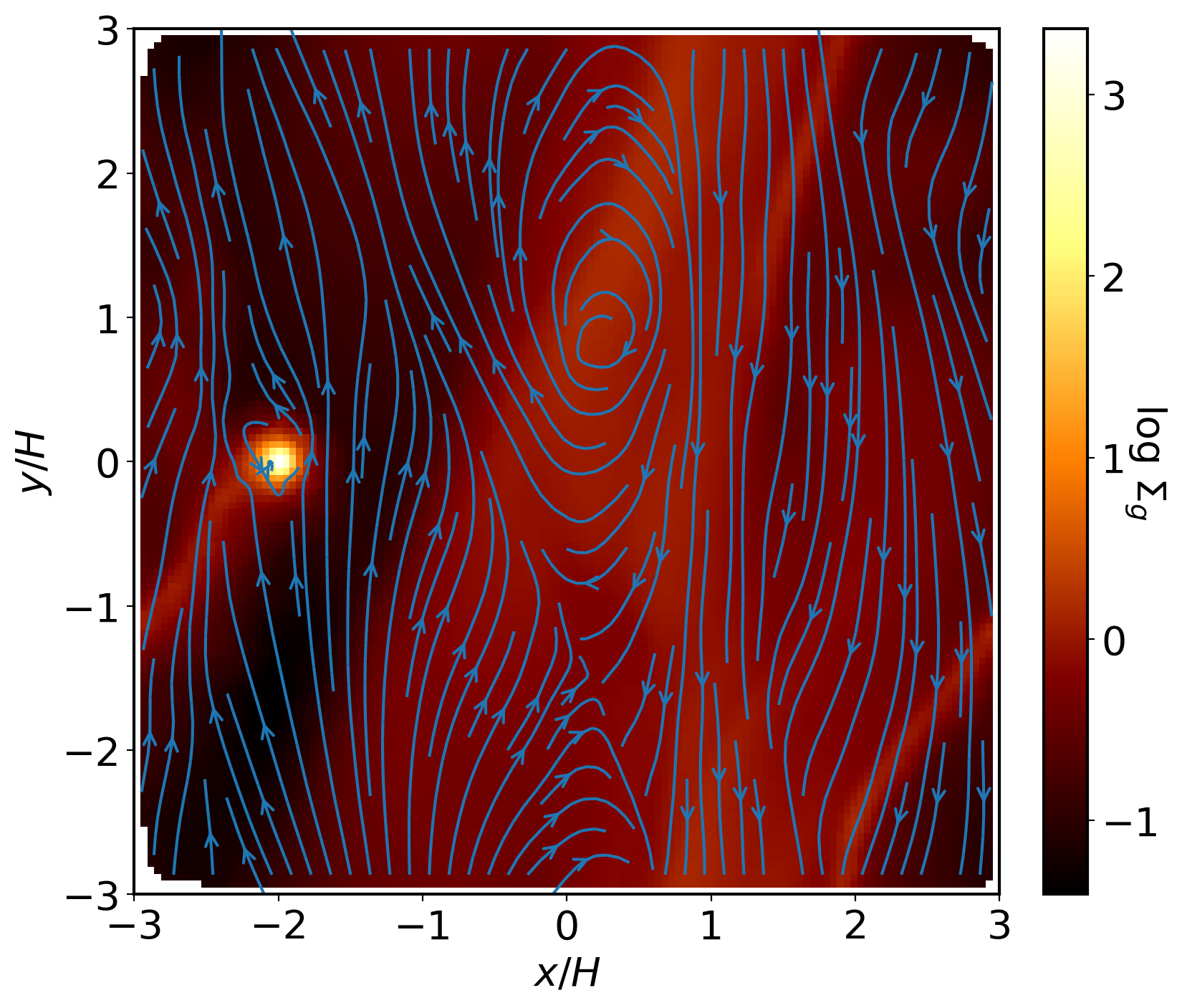}
\caption{Left: radial distribution of the gas pressure, $P/P_0$, for $M_p/M_{\rm th} = 2.25$ at different times. The planet of mass 2.5$M_{\rm th}$ is located at $x = -2H$. Note that a Gaussian bump external to the planet's orbit establishes after a few hundred orbital times. Right: 2D field of the gas density with streamlines of the gas velocity field (blue lines). We observe the formation of vortices which is expected given that the pressure bump created by the planet is formally unstable to Rossby wave instability. All our numerical experiments show the same qualitative behavior.} \label{fig_bump_formation}
\end{figure*}

Figure \ref{fig:limits_tau} illustrates the region of the parameters space $\delta_b-\tilde{w}_{b}$ that produces a pressure bump which meets the trapping requirement outlined above. By fitting a Gaussian function to the planet-induced bump, we find $\delta_b = 1.44$ and $\tilde{w}_b = 0.96$, strong enough to meet our trapping conditions. 
The most widely-variable parameter is therefore $\tau_{s}$. For very large $\tau_{s} > 1$, the arguments above show that {\em no} physically-reasonable values of ``bump'' parameters can actually trap the dust (such grains are decoupled from the gas after all). So we focus on smaller grains, with $\tau_{s} \sim 0.01-1$. We do not explore smaller $\tau_s$ to keep the runtimes of our simulations reasonable.

Note that with our choice of $\Pi = 0.05$ which is approximately the disk aspect ratio, our trapping requirements imply that these bumps may be Rossby-wave unstable \citep[e.g., see][their Table 2, case iv]{Ono16}. \citet{Ono16} provide a fitting formula for the maximum Gaussian amplitude for stability against Rossby wave over two regimes: $0.02 \leq w_b/R \leq 0.05$ and $0.05 \leq w_b/R \leq 0.2$. At the boundary $w_b/R = 0.05$ (equivalent to our $\tilde{w}_b = 1$), the two fitting formula differ by at least an order of magnitude. Nevertheless, according to either of their criteria, our bump ($\delta_b/\tilde{w}_b = 1.5$) is expected to be Rossby-wave unstable. 

\begin{figure*}
\centering
\includegraphics[width=8cm]{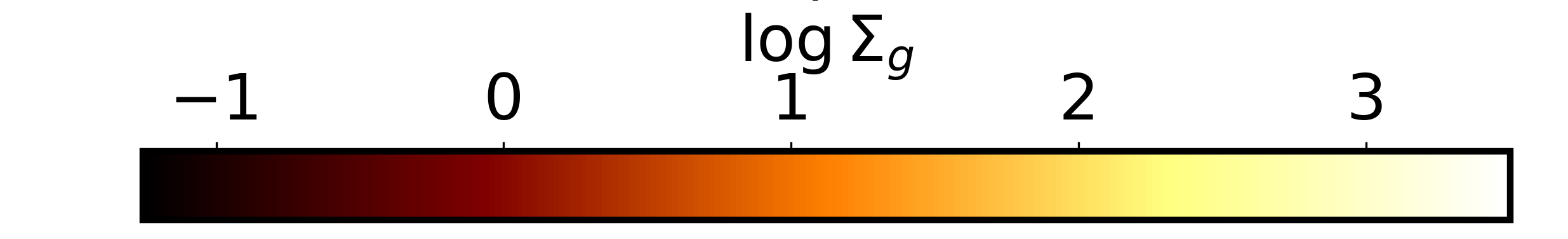}\\ 
\includegraphics[width=\textwidth]{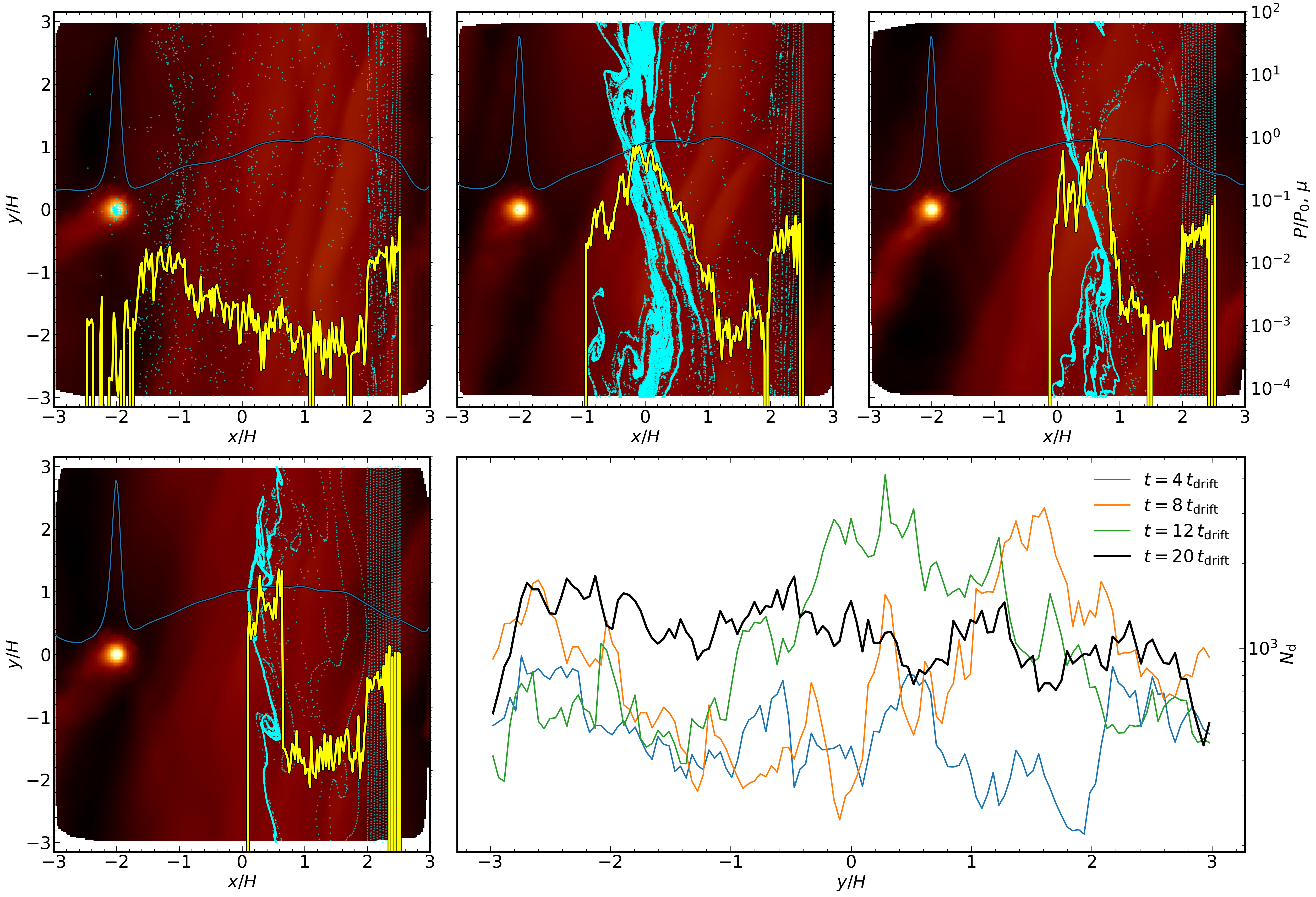}
\caption{Top row + bottom left panel: snapshots of 2D gas density field (background) and the spatial distribution of dust particles (cyan points) for $\tau_s = 0.05$ (top left), 0.10 (top middle), 0.25 (top right), and 0.50 (bottom left), shown here at 5$t_{\rm drift}$. In all cases, the pressure bump was created by a planet of mass $M_p = 2.25M_{\rm th}$, and the planet's location is ($-2H,0$). The solid-lines correspond to radial profiles of the gas pressure ($P/P_0$ in blue) and dust-to-gas mass ratio ($\mu$ in yellow). The snapshots are shown at $t =  5\, t_{\rm drift}$. Bottom right: particle number distribution along the azimuthal coordinate ($y/H$) for $\tau_s=0.1$. While the particles show non-axisymmetric pattern, they evolve towards a more axisymmetric distribution. We observe the same qualitative behavior for particles of larger $\tau_s$.} \label{fig:dust_dritribution_2D_planet}
\end{figure*}

\section{Identifying Dust Rings}\label{sec:planet_results}

\subsection{Gap formation and pressure bump} \label{sec:planet_bump}

With the introduction of gravity from a thermal mass object, some of the surrounding gas is rapidly accreted onto the planet and density waves are excited in the gas, pushing the gas away toward the radial boundaries. This process results in a density gap around the planet's orbit and a pile up of gas a few $H$ away from the planet (see left panel in Figure \ref{fig_bump_formation}). The gas responds to the initial perturbation on a short time and equilibrates after a few hundred orbital times. As expected of bumps strong enough to be Rossby wave unstable, we observe vortices near the bump (see the gas streamlines in the right panel of Figure \ref{fig_bump_formation}). 

We note that after about $\gtrsim$20 orbital times, the gas pressure at the location of the planet remains roughly constant. To make sense of this time scale, we estimate analytically the characteristic time for the accretion process.
For 2D accretion, the gas mass accretion rate is given by
\begin{equation}
\dot{M} \sim R_{\rm acc,g} \Sigma_{\rm g} v_{\rm acc,g}\, ,
\end{equation}
where $R_{\rm acc,g}$ and $v_{\rm acc,g}$ are the accretion radius and velocity, respectively, and $\Sigma_{\rm g}$ is the gas surface density. For $M_p \sim M_{\rm th}$, the Hill radius of the planet becomes smaller than its Bondi radius, and therefore $R_{\rm acc,g}\sim R_{\rm Hill}$.  As the ambient gas flow approaches $R_{\rm acc,g} \sim R_{\rm Hill}$, it will reach the shear velocity: $v_{\rm acc,g} \sim \frac{3}{2}\Omega R_{\rm Hill}$ . The gas accretion rate onto the planet is then 
\begin{equation}
\dot{M} \sim \dfrac{3}{2} R^2_{\rm Hill} \Sigma_{\rm g} \Omega \, .
\end{equation}
As the planet orbits the star, it sweeps up the gas material around on a time scale $t_{\rm sweep} \sim M_{\rm av}/\dot M$, where $M_{\rm av} = 4\pi\Sigma_{\rm g} r R_{\rm Hill}$ is the mass available to the planet at orbital distance $r$.  Noting that $R_{\rm Hill} = f H$, where $f = (2M_p/9M_{\rm th})^{1/3}$, and recalling that for a thin disk $\Pi \sim c_s/U_K \sim H/r$, we obtain
\begin{equation}
t_{\rm sweep} \sim 14\left(\dfrac{M_p}{M_{\rm th}}\right)^{-1/3} \Pi^{-1} \Omega^{-1}\, . \label{eq_tac}
\end{equation}
As a check, for $M_p/M_{\rm th} = 2.25$, and $\Pi = 0.05$, equation \ref{eq_tac} gives $t_{\rm sweep} \sim 207\, \Omega^{-1}$, approximately within an order of magnitude of the time it took for the gas at the location of the planet to reach some steady state in our simulation. The longer $t_{\rm sweep}$ we arrive at likely reflects the difference between a global view adopted in our analytic calculations here compared to the local box approximation in our numerical simulations. 

\subsection{Dust distribution} \label{sec:dist_bump_planet}

Figure \ref{fig:dust_dritribution_2D_planet} visualizes the radial distribution of the gas pressure and dust-to-gas mass ratio, as well as the 2D spatial distribution of dust particles. 
We find that the morphology of the dust band is strongly dependent on $\tau_s$. For $\tau_s = 0.05$, any dust concentration we see is transient and is advected away following the gas flow onto the planet over just one drift time. We also observe more complex geometry of the dust ring with signatures of vortices, likely following the vortices in the gas streamlines (Figure \ref{fig_bump_formation}). In general, the gas pressure bump is constantly deformed not just by the dust feedback but also by the density waves driven by the planet. The complex morphology of gas streamlines begets the complex morphology of dust bands.

For particles of $\tau_s =0.05$, the pressure bump is an ineffective barrier. We observe the particles going through the bump and arriving at the location of the planet, from where they are constantly kicked out of the box by gas outflows. For $\tau_s$ = 0.1, 0.25, and 0.5, we find that particles become trapped slightly inside the center of the bump, as expected for a disk with a smooth pressure gradient on top of a local pressure maximum. We find that $\mu \sim 1$ in these locations and that the radial extent of the dust-rich region becomes smaller for larger $\tau_s$ as larger particles are more strongly affected by aerodynamic drag and able to collect into a pressure maximum more quickly. The fact that the large particles are more decoupled from the gas also implies that they are more resilient against the advective outflow from the gas bump. We note that in spite of the initial vortex formation in dust rings, over time, the dust concentrations transition to axisymmetric rings (see the bottom right panel of Figure \ref{fig:dust_dritribution_2D_planet}).

\subsection{Trap efficiency} \label{sec:eff_bump_planet}

\begin{figure}
\centering
\includegraphics[width=8cm]{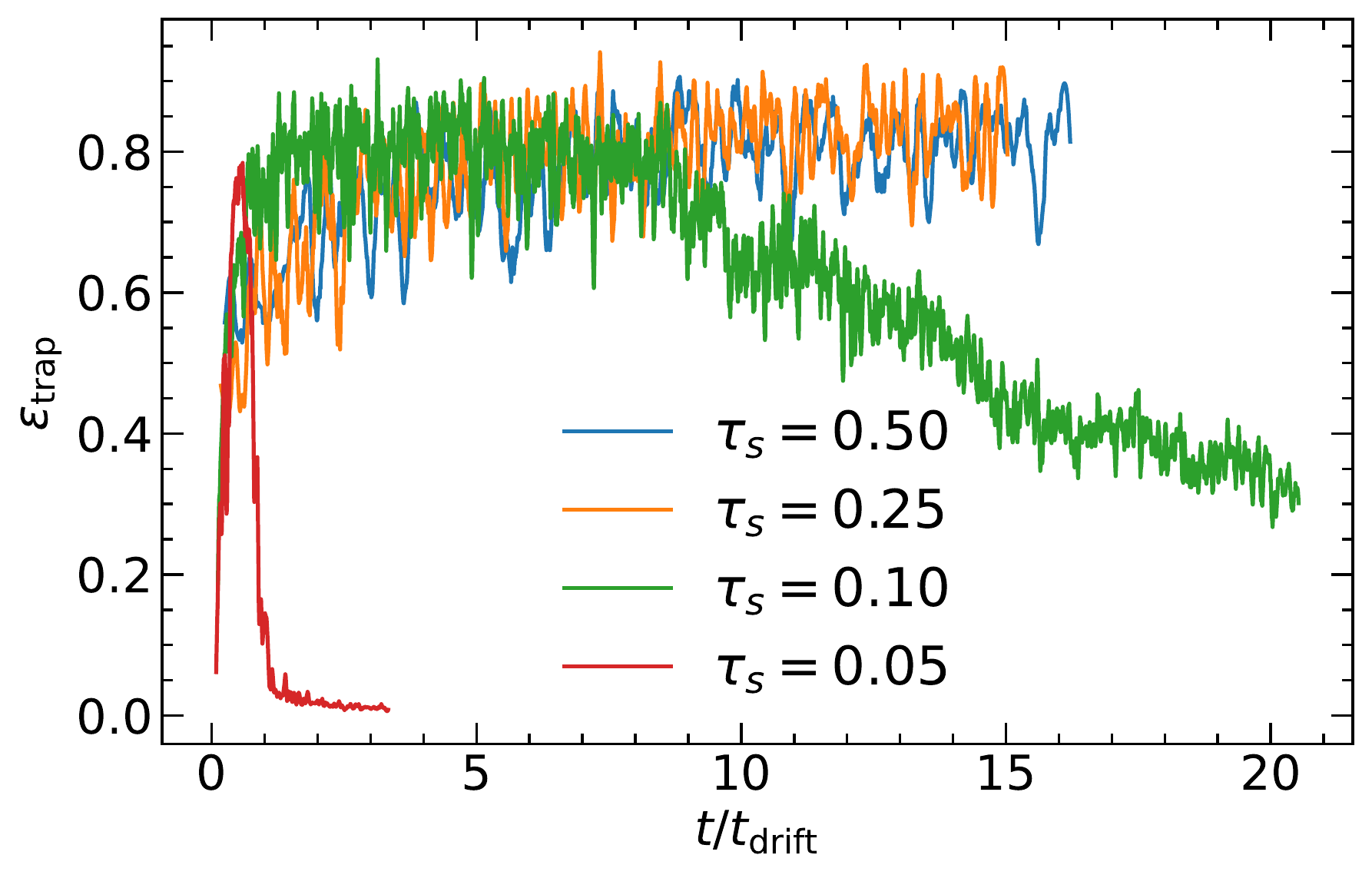}
\caption{Trap efficiency $\epsilon_{\rm trap}$ of planet-driven rings as a function of time. Planet mass is fixed to $M_p = 2.25 M_{\rm th}$. At the lowest $\tau_s$, the dust ring is advected away within $\sim$1 drift time and we also see a gradual loss of particles from the ring of particles at $\tau_s = 0.10$.} \label{fig:eff_planet_bump}
\end{figure}

We calculate the efficiency of the trap $\epsilon_{\rm trap}$ as the ratio between the number of dust particles inside the ``dust-rich bands'' and the cumulative number of particles supplied to the bump at a given time. We define a dust-rich band by fitting a Gaussian function to the radial distribution of dust grains, centered at the peak of such distribution, and the ``total'' width of the ring is taken as two Gaussian standard deviations.

Since we want to compare results for particles of different $\tau_s$, which are supplied to the box at a different rate, in the following we present results at times normalized by the radial drift time across the box
\begin{equation}
t_{\rm drift} \sim \left(\dfrac{N_H}{2 \Pi}\right)\left(\dfrac{1+\tau_s^2}{\tau_s}\right)\Omega^{-1}\, , \label{eq:drift_time}
\end{equation}
where $N_H$ is the number of pressure scale heights ($N_H = 6$ in our simulations). For particles of $\tau_s = (0.05,\,0.1,\, 0.25,\, 0.5)$ , the radial drift times are $t_{\rm drift} \approx (1203,\,606,\, 255,\, 150)\, \Omega^{-1}$, respectively.

As demonstrated in Figure \ref{fig:eff_planet_bump}, $\epsilon_{\rm trap}$ quickly rises to $\sim$0.8 within $\sim$1$t_{\rm drift}$ then either decreases with time or stays constant (at least for the duration of our simulations), depending sensitively on $\tau_s$. We identify the source of the high initial trapping efficiency with vortices acting as effective dust traps. At lower $\tau_s$ (especially for $\tau_s = 0.05$), the particles, being coupled to the gas flow, are eventually advected out of the pressure bump before they can collect into thin rings. We find this behavior persists when we turn off dust feedback---in fact, dust feedback aids the stabilization of dust rings against advection---and we did not observe any noticeable difference in the morphology of planet-induced pressure bump when we increased (or decreased) the resolution from our fiducial $128^2$ particles, suggesting the effect is not dominated by e.g.\ numerical viscosity. For these small $\tau_s$, the dust band leaks out more easily when we turn on planet gravity on dust grains as the grains are attracted to the planet on top of being advected out following the gas flow. As $\tau_s < 0.1$ grains have been shown to collect into thin rings under the presence of a planet both sub- and super-thermal over thousands of orbital times in global disk simulations with different numerical schemes \citep[e.g.,][]{Dong17}, we suspect that the transient ring we observe may be a feature of our local shearing box simulation, which will need to be verified (in the future) using 2D global simulations with GIZMO. 

For particles of $\tau_s= 0.1$, the bump is able to trap dust and $\epsilon_{\rm trap}$ remains high until $t \approx 8\, t_{\rm{drift}}$. Afterwards, we find that $\epsilon_{\rm trap}$ decays with time until the system reaches an equilibrium between the number of particles that escape from the bump and the number of particles supplied to the bump. Finally, for particles of $\tau_s= 0.25$ and 0.5, we do not observe a significant particle leak, and the efficiency of the trap remains high and stable over long time-scales. 

\section{Initial mass reservoir and rings} \label{sec:rings}

\begin{figure}
\centering
\includegraphics[width=8cm]{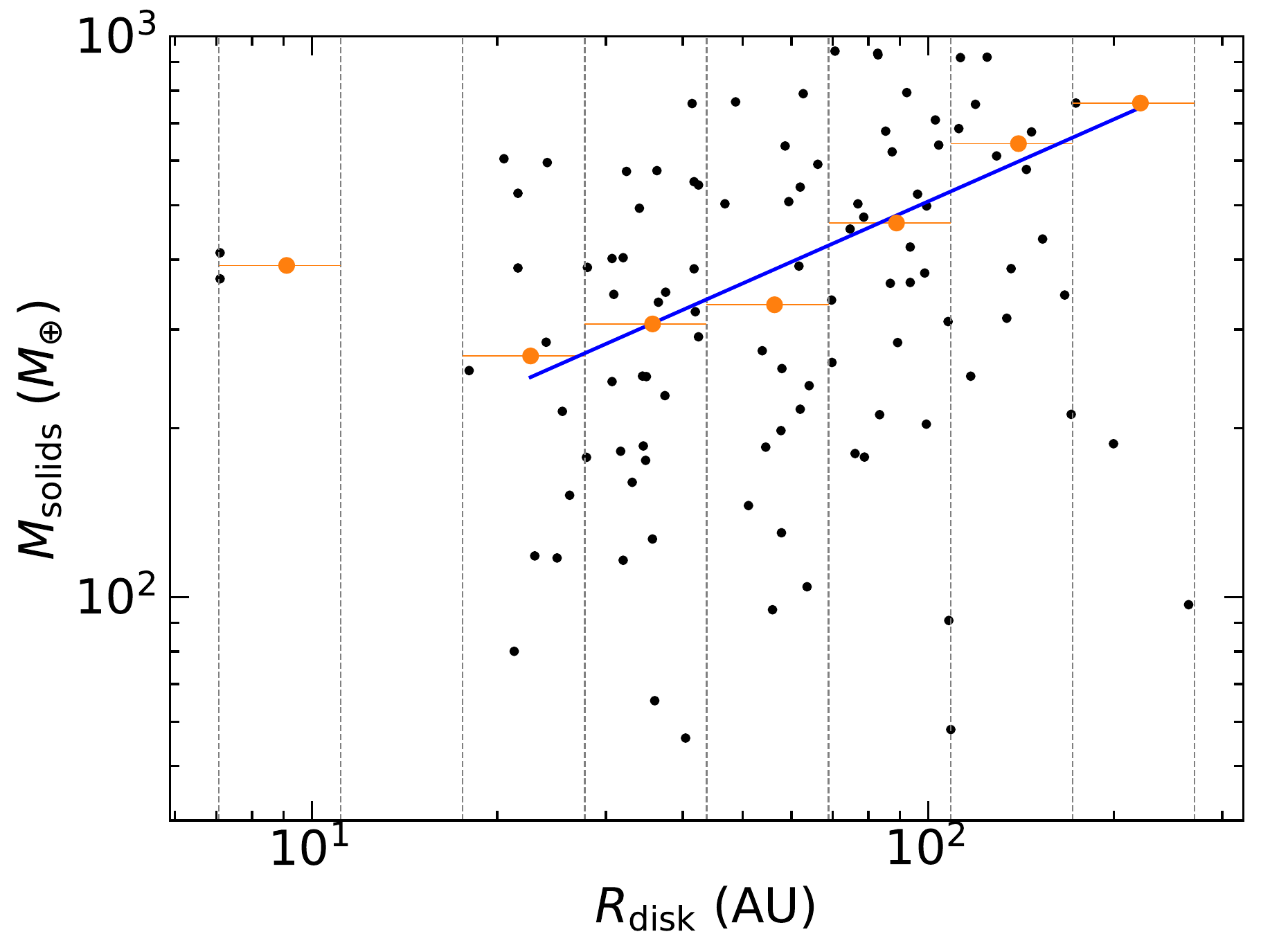}
\caption{Disk dust mass and radius estimates for Class 0 and I sources \citep[][black circles]{2020ApJ...890..130T} where we take the mass measurements from VLA and the size measurement from ALMA (see text for the rationale). We binned the data over $R_{\rm disk}$ (the vertical lines denote the edge of each bin) and take the mean value of each bin as a representative value of $M_{\rm solids}$ (orange dots). The blue line is a power law fit to the last six bins (equation \ref{eq:M_fit}). } \label{fig_masses_vla}
\end{figure}

We now evaluate the amount of solid mass that could be trapped in the rings we simulate and compare to the inferred ring masses in ALMA disks.
The initial solid mass reservoir is inferred from the mass and radius measurements of Class 0/I disks 
in the Orion cluster that are detected with both ALMA (0.87 mm) and the VLA (9 mm), reported by \citet{2020ApJ...890..130T}. 
Following the procedure of \citet{Chachan22}, we take the dust masses from VLA observations as disks are expected to be more optically thin at longer wavelengths (and so they are closer to the true masses). Disk sizes are taken from ALMA measurements as protoplanetary disks tend to appear smaller at longer wavelengths \citep[e.g.,][]{Tazzari16} which may be an effect of different optical depths \citep{Tripahi18}.
By taking the average dust mass of these young disks at each radius bin, we obtain the initial solid mass profile
(see Fig. \ref{fig_masses_vla}):
\footnote{We exclude the first bin since typical ALMA rings are located at orbital distances distances beyond 10-20 AU.} 
\begin{equation}
M_{\rm solids} (R_{\rm disk}) \approx 54 M_{\oplus} \left( \dfrac{R_{\rm disk}}{1\, \mathrm{AU}}\right)^{0.49}\, . \label{eq:M_fit}
\end{equation}

By integrating the radial drift velocity in equation \eqref{eq:vel_drift}, we obtain the initial location from which dust grains are sourced
\begin{equation}
R_0 (t) = R_{\rm f} \left[ 1 + 3\Pi \left(\dfrac{\tau_s}{1 + \tau_s^2}\right) \Omega (R_{\rm f}) \left(\dfrac{H}{R_{\rm f}}\right) t \right]^{2/3}\,  , \label{eq:R_0}
\end{equation}
where $R_{\rm f}$ is the orbital distance of the dust after drift in over a time $t$.
For a dust ring located at $R_{\rm f}$, we use equation \eqref{eq:M_fit} to compute the total dust mass that drifts into $R_{\rm f}$ at any given time:
\begin{equation}
M_{\rm avail} (R_0 (t)) \approx 54 M_{\oplus} \left[\left(\dfrac{R_0 (t)}{1\, \mathrm{AU}}\right)^{0.49} - \left(\dfrac{R_{\rm f}}{1\, \mathrm{AU}}\right)^{0.49} \right]. \label{eq:M_re}
\end{equation}
We can then express the total dust dust mass in the ring at time $t$ as
\begin{equation}
    M_{\rm ring}(t) = \int^t_{t_0} \epsilon_{\rm trap}(t') \left(\frac{dM_{\rm avail}(t')}{dt'}\right) dt',
    \label{eq:m_bump_t}
\end{equation}
where $t_0$ is the time at which we identify a dust ring for each simulation. We stop the integration at time $t$ when $R_o(t) = 200$ AU, taken as the maximum size of a solid disk.

\begin{figure}
\centering
\includegraphics[width=0.5\textwidth]{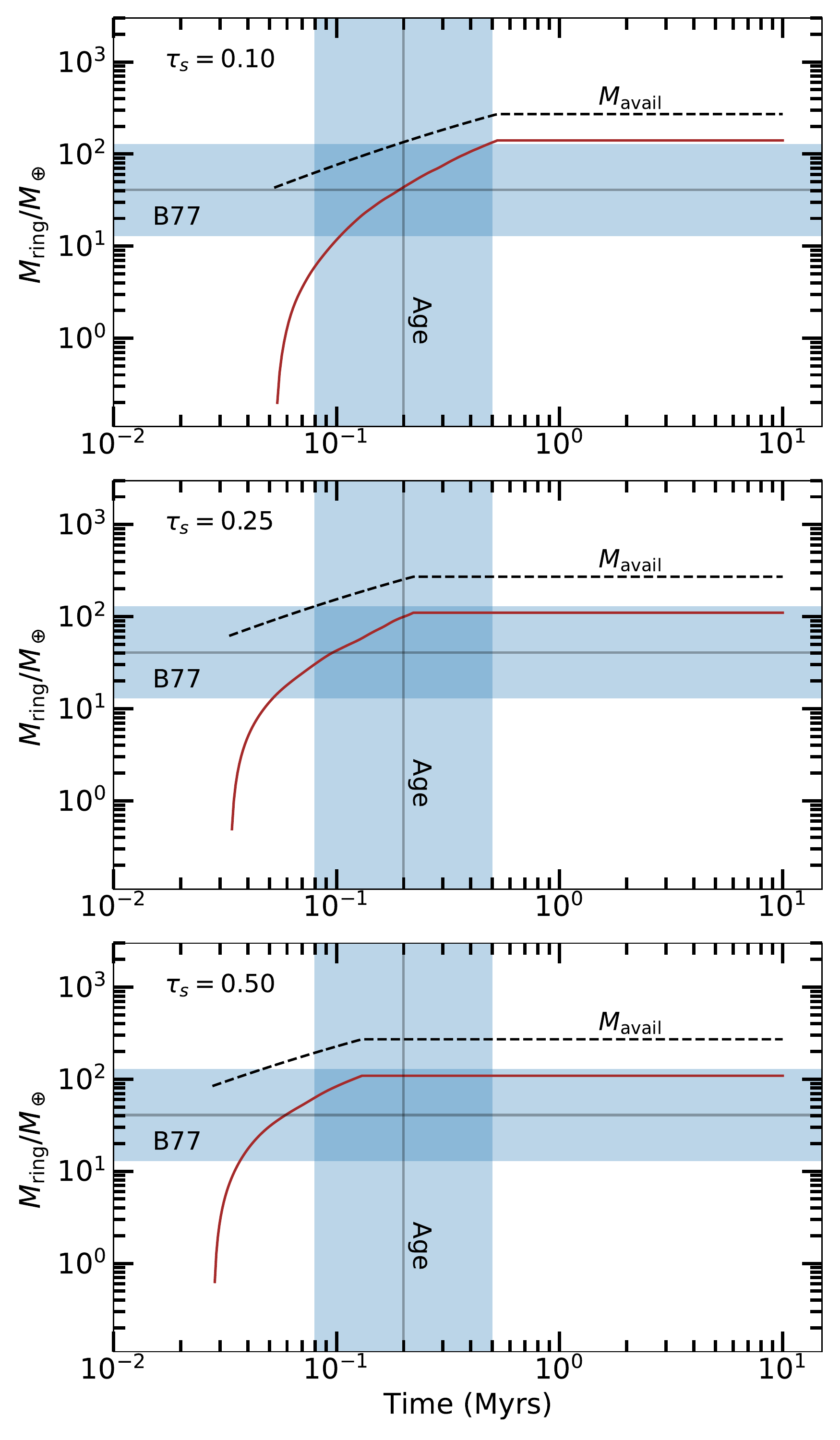}
\caption{Dust mass inside a ring as a function of time (equation \ref{eq:m_bump_t}) for different particle Stokes number $\tau_s$. The total amount of solid mass that would have drifted into the ring is annotated with $M_{\rm avail}$ (see equation \ref{eq:M_re}). The inferred mass of the ring B77 \citep{2018ApJ...869L..46D} and the age of the host system Elias 24 \citep{Andrews18} are represented by the horizontal and vertical lines, respectively, with their 1-$\sigma$ error illustrated with blue bars.} \label{fig:masses_rings}
\end{figure}

For a quantitative comparison with real data, we use the properties of the ring B77 in Elias 24 for its ``median'' properties---i.e., this ring has an approximately median mass and orbital distance out of all the rings studied by \citet{2018ApJ...869L..46D}; furthermore, only a single ring is resolved and so it is more closely analogous to our simulations. The total estimated mass of the ring is $\approx 40.8\, M_{\oplus}$, and it is located at $R_f=76.7$ AU from a star of mass $M_{\star} = 0.78\,M_{\odot}$. Placing the center of our simulation box at $R_f$ and using the $M_{\star}$ of Elias 24, one orbital time in our simulation corresponds to $\sim$760 years and so we scale our simulation times to this value. Figure \ref{fig:masses_rings} demonstrates that all the dust rings in our simulations are able to collect enough mass fast enough to match the inferred mass of B77 and the age of its host system Elias 24 within 1-$\sigma$ uncertainty. (We do not show $\tau_s = 0.05$ case since its dust ring is transient.) The maximum ring mass is reached earlier at larger $\tau_s$ since larger grains undergo more rapid drift.

\subsection{Dust ring evolution}

Another way to compare our simulated dust rings with those observed is to analyze the width of the rings. Assuming the dust rings are established by the drift-diffusion steady state:
\begin{equation}
    \Sigma_{\rm solid}v_x = D_{d,{\rm ring}} d_r \Sigma_{\rm solid}
\end{equation}
where
\begin{equation}
    D_{d,{\rm ring}} \sim v^2_{\rm rms} \frac{\tau_s}{\Omega}
    \label{eq:Dd_ring}
\end{equation}
is the diffusion coefficient of particles inside the ring \citep{2007Icar..192..588Y}, and  
\begin{equation}
v_{\rm rms} = \sqrt{\dfrac{1}{N} \sum_{j}^{N} \left(v_{j,x} - \langle v_x \rangle \right)^2 + \left( v_{j,y} - \langle v_y \rangle \right)^2}\, ,
\label{eq:v_rms_ring}
\end{equation}
is the root-mean-squared dispersion velocity of all particles within the dust ring, defined as $\pm 2\sigma$ from the centre of the Gaussian fit. Here, $j$ and $N$ denote the jth-particle and the total number of dust particles in the dust ring, respectively, and $\langle  \rangle$ is the average of the $N$ particles in the clump. 

While we directly compute $v_{\rm rms}$ numerically, we can express it in terms of gas sound speed:
\begin{equation}
    v_{\rm rms} = c_s\sqrt{\frac{\alpha_{\rm eff}}{1+\tau_s}}
    \label{eq:v_rms_d}
\end{equation}
where $\alpha_{\rm eff}$ is the effective turbulence parameter. We stress that this $\alpha_{\rm eff}$ is limited to `turbulence' within the radial-azimuthal plane and is distinct from the degree of vertical turbulence. For all our simulations, $v_{\rm rms}$ rises with time and $\alpha_{\rm eff} \sim 0.01$--0.1 with larger $\tau_{s}$ characterized by smaller $v_{\rm rms}$ due to their relative ease with being collected into a pressure bump.

\begin{figure}
    \centering
    \includegraphics[width=0.5\textwidth]{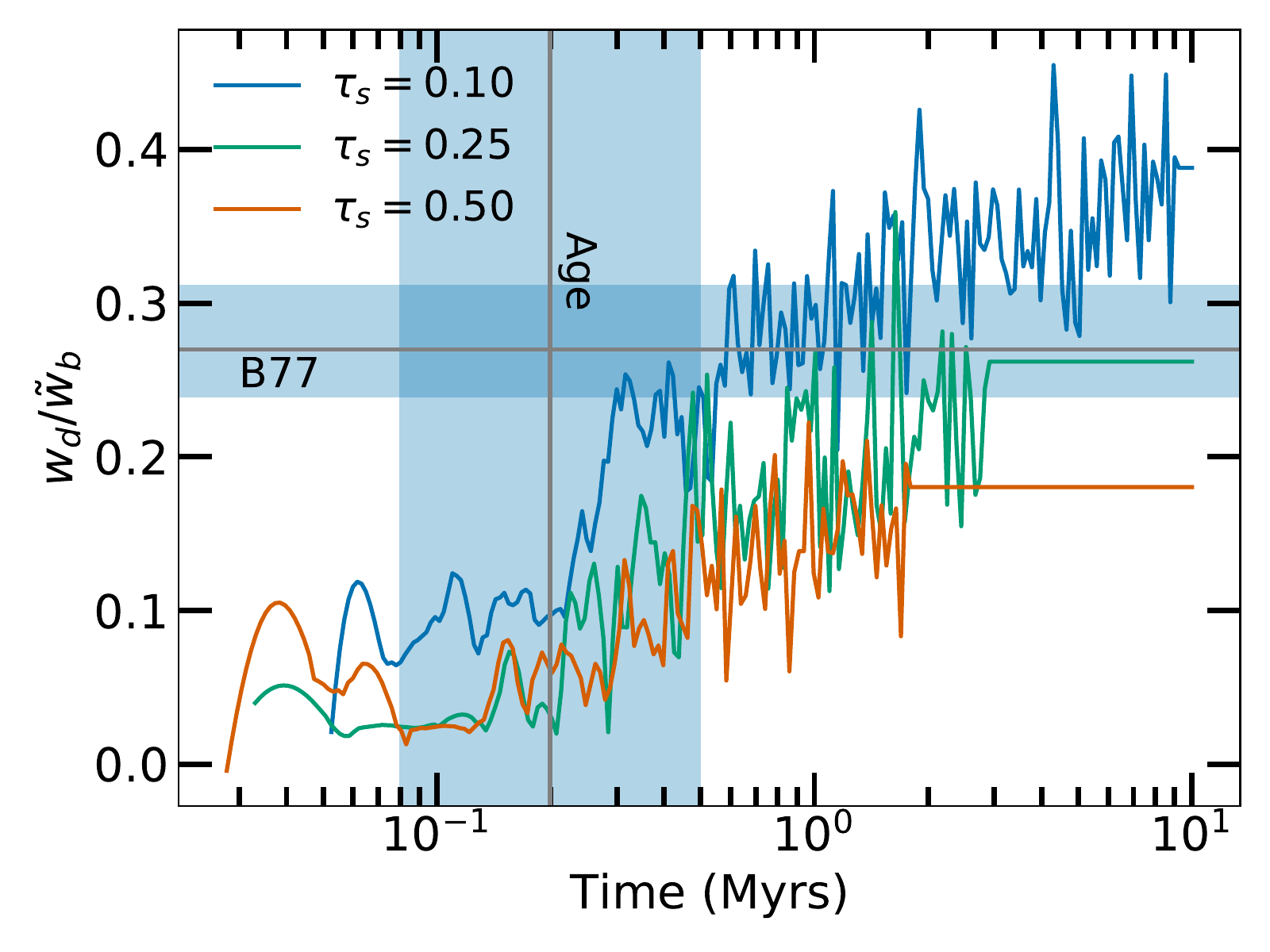}
    \caption{Time evolution of dust ring width in units of the width of the gas pressure bump (set to one 0.96$H$ in all cases). The horizontal and vertical blue bars represent 1-$\sigma$ uncertainty in the ratio of dust-to-gas pressure bump width \citep[][see their Figure 5 and Table 3]{2018ApJ...869L..46D} and in the estimated age of the host system Elias 24 \citep{Andrews18}, respectively. Under planet's tidal forcing, the particle ring puffs up over time.}
    \label{fig:dust_wd_time}
\end{figure}

\begin{figure}
    \centering
    \includegraphics[width=0.5\textwidth]{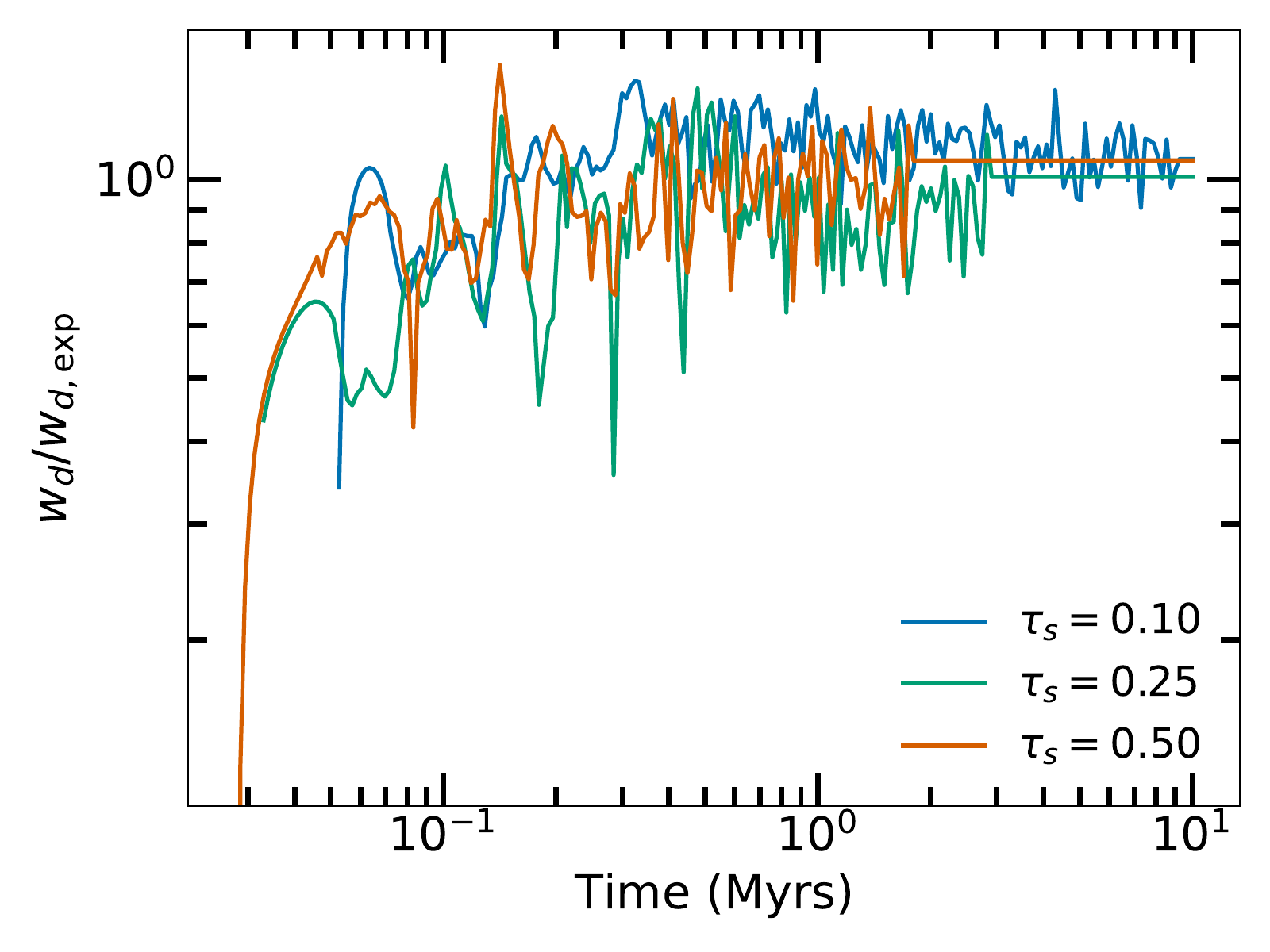}
    \caption{The ratio between the measured dust ring width and the expected width under drift-diffusion steady state (equation \ref{eq:w_d_exp}). The dust ring driven by planet's tidal forcing is described well by the balance between the radial drift and turbulent diffusion.}
    \label{fig:dust_wd_comp}
\end{figure}

If the center of the dust ring is located at the center of the gas pressure bump (in Gaussian form), then $\Sigma_{\rm solid}$ can be expressed as a Gaussian with a width (c.f. equation 46 of \citet{2018ApJ...869L..46D} with $D_{d,{\rm ring}}$ in equation \ref{eq:Dd_ring})
\begin{equation}
    w_{d,{\rm exp}} \sim w_b \frac{v_{\rm rms}}{c_s}\sqrt{(1+\tau^2)} \sim w_b \sqrt{\frac{\alpha_{\rm eff}}{1+\tau}}
    \label{eq:w_d_exp}
\end{equation}
in the limit of $\tau_s < 1$ with $v_{\rm rms}$ expressed as equation \ref{eq:v_rms_d} and $w_b$ representing the width of the gas pressure bump. We then expect the dust to thicken with time as $v_{\rm rms}$ rises as we see in Figure \ref{fig:dust_wd_time}. Figure \ref{fig:dust_wd_comp} demonstrates that the measured width of our dust ring under tidal forcing by a planet tracks well the expected ring width from drift-diffusion steady state.

\begin{figure}
    \centering
    \includegraphics[width=0.5\textwidth]{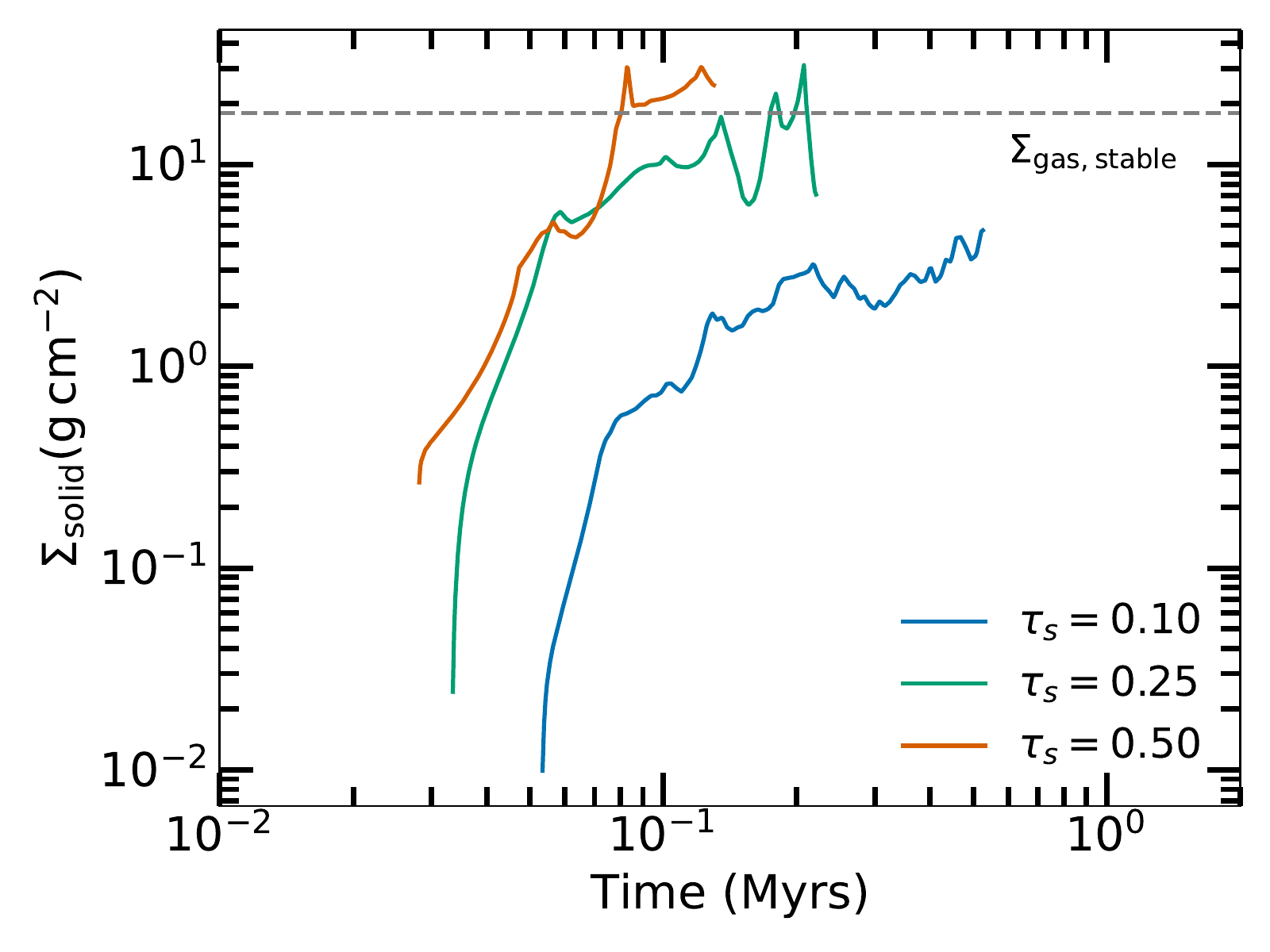}
    \caption{Time evolution of ring solid surface density. Greater ring mass and thinner ring begets larger solid density for larger $\tau_s$. The horizontal dashed line delineates the maximum gas surface density to be Toomre-stable under the assumption of irradiation-dominated midplane temperature (see Table 2 of \citet{2018ApJ...869L..46D}). Values are shown until the solids out to the maximum disk radius 200 AU drift into $R_f=76.7$ AU (see equation \ref{eq:m_bump_t} and the surrounding text).}
    \label{fig:Sig_sol_v_t}
\end{figure}

From Figure \ref{fig:dust_wd_time}, we infer that under tidal forcing by a planet, $\tau_{s}=0.10$ rings can reproduce the width of the B77 ring within the age of Elias 24. Larger $\tau_{s}$ particles tend to create sharper rings as they are more efficiently dragged and collect more easily into pressure traps. Overall, compared to B77, rings of $\tau_{s}=0.25$ and 0.50 have generally larger mass than the median measured value (see Figure \ref{fig:masses_rings}) and are generally thinner than the median quoted width (see Figure \ref{fig:dust_wd_time}), resulting in dense rings. In fact, within $\sim$0.2 Myrs, these high-$\tau_{s}$ rings reach solid surface density $\Sigma_{\rm solid}$ that is comparable to and slightly larger than the maximum gas surface density to be stable against gas self-gravity (estimated under the assumption of irradiation-dominated midplane temperature; see Figure \ref{fig:Sig_sol_v_t}). While this is technically an allowed solution as the local dust-to-gas ratio in $\tau_{s}=0.25$ and 0.50 runs reach $\gtrsim$2--3, and so the local gas density can be smaller than the local solid density, it is still uncomfortably close to the limit of stability. We conclude that the rings observed in the DSHARP survey \citep{Andrews18} can be created by planetary perturbers with the additional constraint that $\tau_{s}$ is more likely $\lesssim 0.10$.

\subsection{Gravitational collapse of dust inside the trap} \label{ssec:grav_collapse_gaussian}

While we do not explicitly simulate the effect of self-gravity of the dust particles in our calculations,\footnote{Gas and dust self-gravity are available only in 3-dimensional version of GIZMO, which is beyond the scope of this manuscript and is a subject of future work.} we can estimate the mass of clumps in the dust-rich bands that are expected to collapse into planetesimals or planetary bodies. We use the virial parameter for a spherical clump \citep[e.g.,][]{1992ApJ...395..140B} to determine the size and mass of the bound clump in our simulations:
\begin{equation}
\alpha_{\rm vir} \sim \dfrac{5v_{\rm rms,cl}^2 R_{\rm cl}}{G M_{\rm cl}} \leq 1 , \label{eq_alpha}
\end{equation}
where $v_{\rm rms,cl}$ is the root mean squared dispersion velocity of the dust particles in the clump, $R_{\rm cl}$ is the clump's radius, and $M_{\rm cl}$ is the total mass of dust in the clump. If $\alpha_{\rm vir} > 1$, dust particles have enough kinetic energy to expand and move through the gas, whereas dust clumps with $\alpha_{\rm vir} \leq 1$ are gravitationally bound. We note that this collapse condition is equivalent (within a numerical factor) to the diffusion-limited collapse criterion for planetesimals outlined by \citet{Klahr18} and \citet{Gerbig20} which derives from the condition where the contraction timescale
\begin{equation}
    t_{\rm contr} = \frac{\Omega}{4\pi G \rho_{\rm cl}\tau_s}
\end{equation}
(where $\rho_{\rm cl} = 3 M_{\rm cl}/4\pi R_{\rm cl}^3$ is the density of the clump)
is shorter than the diffusion timescale
\begin{equation}
    t_{\rm diff} = \frac{R_{\rm cl}^2}{D_d}
\end{equation}
with
\begin{equation}
    D_d \sim v^2_{\rm rms,cl} \frac{\tau_s}{\Omega}
\end{equation}
the particle diffusion coefficient \citep{2007Icar..192..588Y}. The collapse criterion $t_{\rm contr} < t_{\rm diff}$ boils down to
\begin{equation}
    \frac{v^2_{\rm rms,cl} R_{\rm cl}}{3G M_{\rm cl}} \sim \alpha_{\rm vir}/15 < 1.
    \label{eq:collapse_crit}
\end{equation}
By defining a bound clump as those with $\alpha_{\rm vir} < 1$, our dust clumps are guaranteed to collapse against turbulent diffusion.

\begin{figure}
\centering
\includegraphics[width=0.5\textwidth]{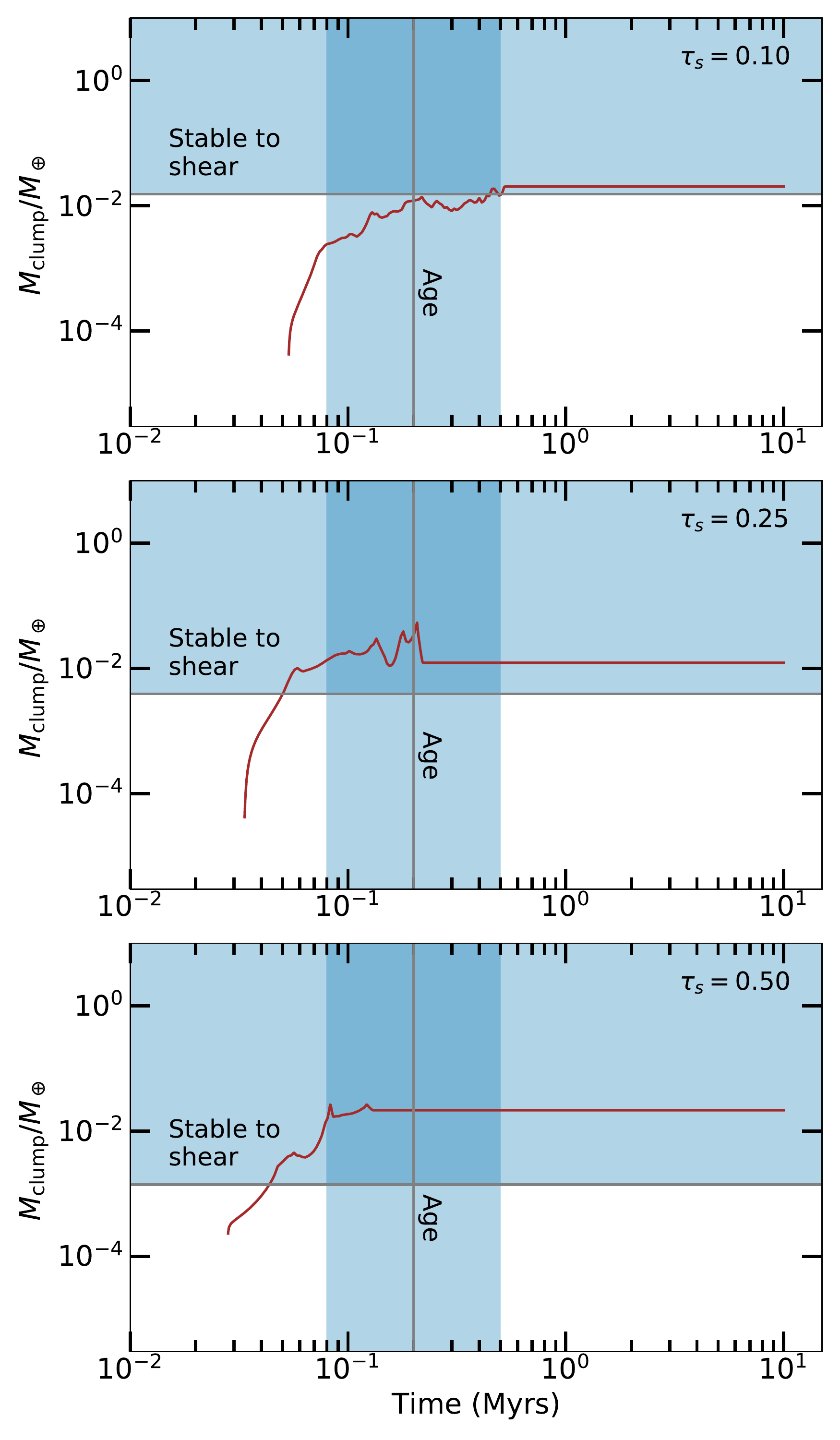}
\caption{Mass evolution of the bound dust clump in the dust-rich bands; annotations are identical to that of Figure \ref{fig:masses_rings}, with the addition of a grey horizontal line delineating the minimum mass for stability against tidal shear. At lower $\tau_{s}$, the bound clumps, at their maximal size, are more likely to be sheared apart.}\label{fig:mass_clumps}
\end{figure}

We identify a bound clump for all our simulations as follows:
\begin{enumerate}
\item We obtain the (x,y)-coordinate of the densest point within the dust ring, defined as a radial strip of total width $4\sigma$ centered at the peak of the Gaussian fit. Before identifying the densest point, we smooth the 2-dimensional distribution of dust grains using a Gaussian kernel density estimator (KDE).\footnote{The smoothness of the KDE is set by the bandwidth parameter, which for the present work is given by the ``Scott’s Rule''\citep[i.e., $N^{-1/(d+4)}$, where $N$ is the number of data points and $d$ the number of dimensions, see, e.g.,][]{scott2015multivariate}.}
This (x,y)-coordinate locates the center of our clump.

\item We first set $R_{\rm cl} = H_{\rm solid}$, where, following \citet{2007Icar..192..588Y}, we define $H_{\rm solid}$ as the dust scale height
\begin{align}
H_{\rm solid} = H\sqrt{\dfrac{\alpha}{\alpha + \tau_s}}\,
\end{align}
with Shakura-Sunyaev parameter $\alpha = 10^{-3}$ as this is the maximum expected value from the geometry of ringed disks \citep{Pinte16} and from CO line measurements in protoplanetary disks \citep[e.g.,][]{Flaherty17}. We note that this $\alpha$ probes the vertical turbulence and is distinct from $\alpha_{\rm eff}$ previously defined.

\item Within a radius of $R_{\rm cl}$ centered at the densest point, we calculate the $v_{\rm rms,cl}$ of dust particles using equation \ref{eq:v_rms_ring}.

\item The mass of the clump is calculated as $M_{\rm cl} = \pi R_{\rm cl}^2 \Sigma_{\rm solid}$, where $\Sigma_{\rm solid}$ is the solid surface density of the dust ring. To estimate $\Sigma_{\rm solid}$ as a function of time, we integrate the Gaussian fit of the dust ring within $2\sigma$, and normalize it to the mass of the ring as computed in equation \ref{eq:m_bump_t}:
\begin{equation}
\Sigma_{\rm solid} \sim \dfrac{M_{\rm ring}}{2\pi C}\, ,
\end{equation}
where
\begin{equation}
C = \int_{R_f- 2\sigma}^{R_f + 2\sigma } e^{-0.5(x-R_f)^2/\sigma^2} x\, dx\,.
\end{equation}
The normalization $C$ is close to and slightly larger than $2 R_f \sigma$. We note that in all our simulation snapshots (except at the very earliest times), the x-coordinate of the center of the clump is close enough to the radial center of the dust ring so that we do not need to worry about the Gaussian fall-off in $\Sigma_{\rm solid}$.

\item If the clump's virial parameter at $R_{\rm cl} = H_{\rm solid}$ is larger than 1, we iteratively shrink $R_{\rm cl}$ and follow steps 3 and 4 above until we reach $\alpha_{\rm vir} \leq 1$. We find that step 5 is never invoked in any of our simulation snapshots (i.e., all our ``bound'' clumps are at the maximum size).

\end{enumerate}

In protoplanetary disks, clumps that can gravitationally collapse against turbulent diffusion may still be sheared apart. To be stable against tidal shear, the clump's self-gravity must be larger than tidal acceleration in 3-body dynamics \citep{Gerbig20}:
\begin{equation}
    \frac{GM_{\rm cl}}{R_{\rm cl}^2} > 3\frac{GM_\star}{R_f^2}\left(\frac{R_{\rm cl}}{R_f}\right).
\end{equation}
We find that $R_{\rm clump} = H_{\rm solid}$ always to keep the clump's $\alpha_{\rm vir} \leq 1$ and so setting the clump radius as the dust scale height, the condition for stability against shear:
\begin{equation}
    M_{\rm cl} > 3 M_\star \left(\frac{H}{R_f}\right)^3 \left(\frac{\alpha}{\alpha+\tau}\right)^{3/2}.
    \label{eq:Mclump_stable}
\end{equation}

As illustrated in Figure \ref{fig:mass_clumps}, our dust rings, scaled to the properties of B77 in Elias 24, are able to nucleate stable bound clumps although the $\tau_{s}=0.10$ ring is expected to nucleate clumps that are just barely massive enough to be stable against tidal shear. The increasing difficulty in creating stable planetesimal/planetary bodies at lower $\tau_{s}$ stems from two effects. At small $\tau_{s}$, the minimum clump mass to be stable against shear is larger because of larger $H_d$ (i.e., clumps are more extended). Dust rings need to collect more mass to reach the stability limit but this collection takes a while since the radial drift is slower at smaller $\tau_{s}$ so that at a given time (i.e., the given age of the system), $M_{\rm ring}$ is smaller. Furthermore, rings made of small $\tau_{s}$ particles are puffier and so $\Sigma_{\rm solid}$ drops even more, reducing $M_{\rm cl}$.
We note that creating a stable clump within the age of the system becomes easier even at small $\tau_s$ if the dust ring is located closer to the star where the dynamical timescales are shorter and if the system is older. 

At large $\tau_s$, the minimum clump mass for shear-stability is smaller because of smaller $H_d$ (i.e., clumps are more compact) and so dust rings do not need to collect as much mass. Nevertheless, larger $\tau_{s}$ rings tend to create more massive clumps since their rings are narrower (see equation \ref{eq:w_d_exp}), and so $\Sigma_{\rm solid}$ is boosted. With the innate ability to nucleate more massive clumps and with the minimum mass for stability lower, it is significantly easier to maintain these bound clumps with larger $\tau_{s}$ particles. 

\begin{figure}
    \centering
    \includegraphics[width=0.5\textwidth]{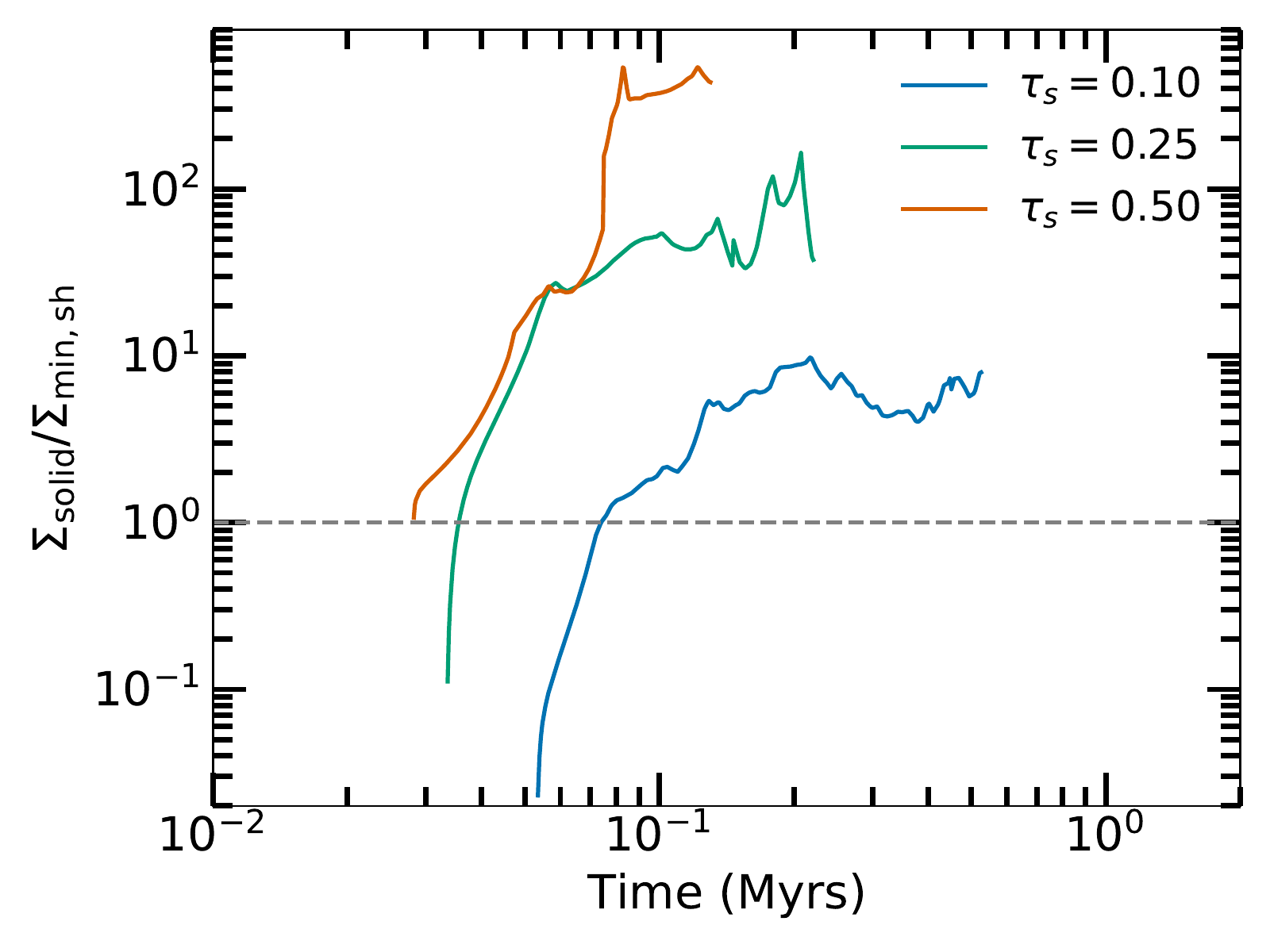}
    \caption{The ratio of solid density in the dust ring to the minimum density required to for a clump to be bound against turbulent diffusion and be stable against tidal shear (see equation \ref{eq:sigsol_crit}). Values are shown until the solids out to the maximum disk radius 200 AU drift into $R_f=76.7$ AU. Except for the initial fraction of evolution, all dust rings are dense enough to nucleate smallest stable planetesimals.}
    \label{fig:Sig_sol_shear}
\end{figure}

It may be possible that we could define a smaller dust clump $R_{\rm cl} < H_{\rm solid}$ so that it is shear-stable within the age of Elias 24 for any $\tau_{s}$. Consider $R_{\rm cl} = f_R H_{\rm solid}$ where $f_R \leq 1$ is a numerical factor. The collapse criterion against turbulent diffusion (equation \ref{eq:collapse_crit}) sets the lower limit on $f_R$:
\begin{equation}
    f_R > \frac{v^2_{\rm rms,cl}}{3\pi G\Sigma_{\rm solid} H_{\rm solid}} \equiv f_{R,{\rm min}},  
    \label{eq:fR_crit}
\end{equation}
where we used $M_{\rm cl} = \pi \Sigma_{\rm solid} R_{\rm cl}^2$. In order for this clump to be stable against tidal shear, 
\begin{equation}
    \frac{v^2_{\rm rms,cl}}{3\pi G\Sigma_{\rm solid}} < (R_{\rm cl} = f_R H_{\rm solid}) < \frac{\pi \Sigma_{\rm solid} R_f^3}{3M_\star}.
\end{equation}
It follows that this condition will be met if
\begin{equation}
    \Sigma_{\rm solid}^2 > \frac{M_\star v^2_{\rm rms,cl}}{\pi^2 G R^2_f} \equiv \Sigma_{\rm min, sh}^2.
    \label{eq:sigsol_crit}
\end{equation}

\begin{figure}
    \centering
    \includegraphics[width=0.5\textwidth]{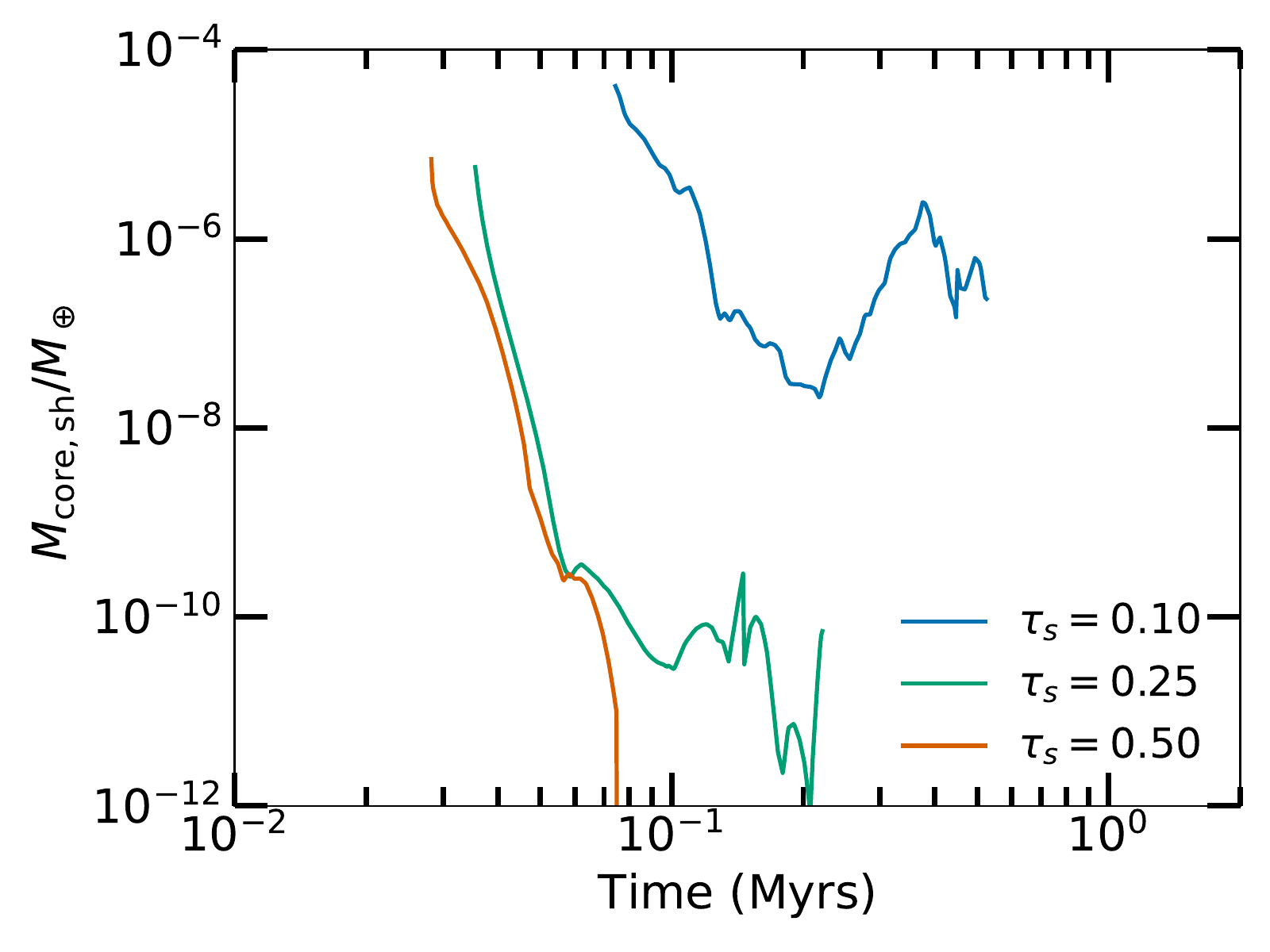}
    \caption{Masses of smallest possible bound clumps stable against tidal shear. Values are shown from when $\Sigma_{\rm solid} \geq \Sigma_{\rm min, sh}$ and until the solids out to maximum disk radius 200 AU drift into $R_f=76.7$ AU.}
    \label{fig:minMcore_shear}
\end{figure}

As demonstrated in Figure \ref{fig:Sig_sol_shear}, for the majority of the evolution, our dust rings meet the density criterion for the creation of the smallest planetesimal stable against tidal shear. Again, we observe that satisfying the stability criterion against shear is increasingly harder for smaller $\tau_{s}$ due to their larger $v_{\rm rms,cl}$ (and therefore larger $\Sigma_{\rm min,sh}$) and smaller $\Sigma_{\rm solid}$ at a given time due to slower radial drift. The corresponding mass of the smallest bound clump stable against shear is
\begin{equation}
    M_{\rm core,sh} = 3M_\star f_{\rm R,min}^3 \left(\frac{H}{R_f}\right)^3 \left(\frac{\alpha}{\alpha+\tau}\right)^{3/2},
\end{equation}
where $f_{\rm R,min}$ is given by the right hand side of equation \ref{eq:fR_crit}. Figure \ref{fig:minMcore_shear} shows that these minimum core masses are smaller than Ceres $\sim$10$^{-4} M_\oplus$.
We conclude that down to $\tau_{s} \sim 0.1$, it is possible to create large planetesimals out to the size of the dust scale height in the dust rings we simulate, and for smaller $\tau_{s}$ (if we can keep these dust rings stable against advection), it is possible to create smaller bodies down to sub-Ceres masses.

\section{Planet formation in dust rings} \label{sec:clump_evol_acc}

Tidal forcing by a planet creates a long-lasting pressure bump that can collect particles into an axisymmetric or nearly axisymmetric ring. Scaled to the properties of B77 in Elias 24 \citep{Andrews18}, we find that all of our rings are able to collect enough dust grains to match the measured mass \citep{2018ApJ...869L..46D}. In addition, all our dust rings are expected to nucleate small, bound and shear-stable clumps. In this section, we investigate the expected mass growth of these clumps via pebble accretion. 

In general, the mass growth rate of a core embedded in a disk of solids is
\begin{equation}
    \dot{M}_{\rm core} = 2\Sigma_{\rm solid}R_{\rm acc}v_{\rm acc} \times {\rm min}(1, R_{\rm acc}/H_{\rm solid})
    \label{eq:Mgrowth_peb}
\end{equation}
where particles that enter within a radius $R_{\rm acc}$ of the core at speeds of $v_{\rm acc}$ will be accreted to the core. Growth by pebble accretion begins in earnest when $\tau < 1$ and when the particle stopping time is shorter than its interaction time with the core (i.e., the ``settling'' regime as identified by \citet{Ormel10}; see also review by \citet{Ormel17}):
\begin{equation}
    \tau_s/\Omega < R_{\rm acc}/V_{\rm acc}.  
    \label{eq:settl}
\end{equation}

We first establish the bound clump mass at which accretion is in this settling regime. Following the procedure of \citet{Lin18}, we compute $R_{\rm acc}$ and $v_{\rm acc}$ in the settling regime and verify that equation \ref{eq:settl} is satisfied. For all our simulations, $\tau_s < 1$ so we use
\begin{equation}
    v_{\rm acc} = \sqrt{\left(v_{\rm hw} + \frac{3}{2}\Omega R_{\rm acc}\right)^2 + v^2_{\rm rms}}
    \label{eq:vacc_comp}
\end{equation}
where $v_{\rm hw} \equiv - (c^2_s/2\Omega a)(\partial\log P/\partial\log a)$ which we compute directly from our simulations as
$-\langle v_x\rangle_\phi (1 + \tau_s^2)/\tau_s$ evaluated at the location of the bound clump (i.e., the radial center of the dust ring) with $<>_\phi$ denoting azimuthal average, $c_s$ the sound speed, $a$ the orbital distance, $P$ the gas pressure, and $v_{\rm rms}$ the rms velocity computed within the dust ring as defined in equation \ref{eq:v_rms_ring}. Under the settling condition (equation \ref{eq:settl}), particles that accrete onto the core attain a terminal velocity during the encounter so that
\begin{equation}
    \frac{v_{\rm acc}}{4} = \frac{GM_{\rm core}}{R^2_{\rm acc}}\frac{\tau_s}{\Omega}
    \label{eq:vacc_settl}
\end{equation}
where $G$ is the gravitational constant. From this, $R_{\rm acc}$ is solved for by finding the root of
\begin{equation}
    \frac{9}{4}b^6 + 3\zeta b^5 + \left(\zeta^2 + \zeta^2_{\rm rms}\right)b^4 - 144\tau^2_s = 0
    \label{eq:bacc}
\end{equation}
where $b \equiv R_{\rm acc}/R_{\rm Hill}$, $R_{\rm Hill} = \mu^{1/3}_M a$, $\mu_M \equiv M_{\rm core}/3M_\star$, $\zeta \equiv v_{\rm hw}/v_{\rm Hill}$, $v_{\rm Hill} = \Omega R_{\rm Hill}$, and $\zeta_{\rm rms} \equiv v_{\rm rms}/v_{\rm Hill}$. We find that dust clumps need to be at least 0.03, 0.03, and 0.1$M_\oplus$ for $\tau_s =$0.1, 0.25, and 0.5, respectively, to be in the settling regime.

For these initial cores to be stable against tidal shear, the dust ring needs to be sufficiently dense. Labeling the minimum core mass for pebble accretion as $M_{\rm settl}$ and letting $M_{\rm settl} = \pi\Sigma_{\rm solid} (f_{R,{\rm settl}} H_{\rm solid})^2$ with $f_{R,{\rm settl}} < 1$, the shear-stability condition can be re-written as
\begin{equation}
    \Sigma_{\rm solid} > \left(\frac{M_{\rm settl} M^2_\star}{\pi^3 R_f^6}\right)^{1/3}.
    \label{eq:settl_cond1}
\end{equation}
In addition,
\begin{equation}
    \frac{M_{\rm settl}}{M_\star} < \left(\frac{H_{\rm solid}}{R_f}\right)^{1/3}
    \label{eq:settl_cond2}
\end{equation}
to ensure $f_{R,{\rm settl}} < 1$. We find that the above conditions are met for all our simulated rings.
%only when $\tau_s = 0.05$ and 0.10 for Gaussian-forcing. \textbf{For larger $\tau_s$, the clump mass required for pebble accretion in the settling regime $M_{\rm settl}$ is too large while $\Sigma_{\rm solid}$ of the ring is too small owing to initially wider ring widths and smaller $\epsilon_{\rm trap}$ and $f_{R,{\rm settl}}$ exceeds unity.} For planet-driven bumps, the above conditions are met for all possible $\tau_s = 0.1$, 0.25, 0.5. \textbf{While $M_{\rm settl}$ is the same between planet-driven and Gaussian-forced bumps for $\tau_s = 0.25$ and 0.5, $\Sigma_{\rm solid}$ is significantly larger in planet-driven bumps owing to initially tighter ring widths and overall larger $\epsilon_{\rm trap}$ which allows for $f_{R,{\rm settl}} < 1$.}

\begin{figure}
    \centering
    \includegraphics[width=0.5\textwidth]{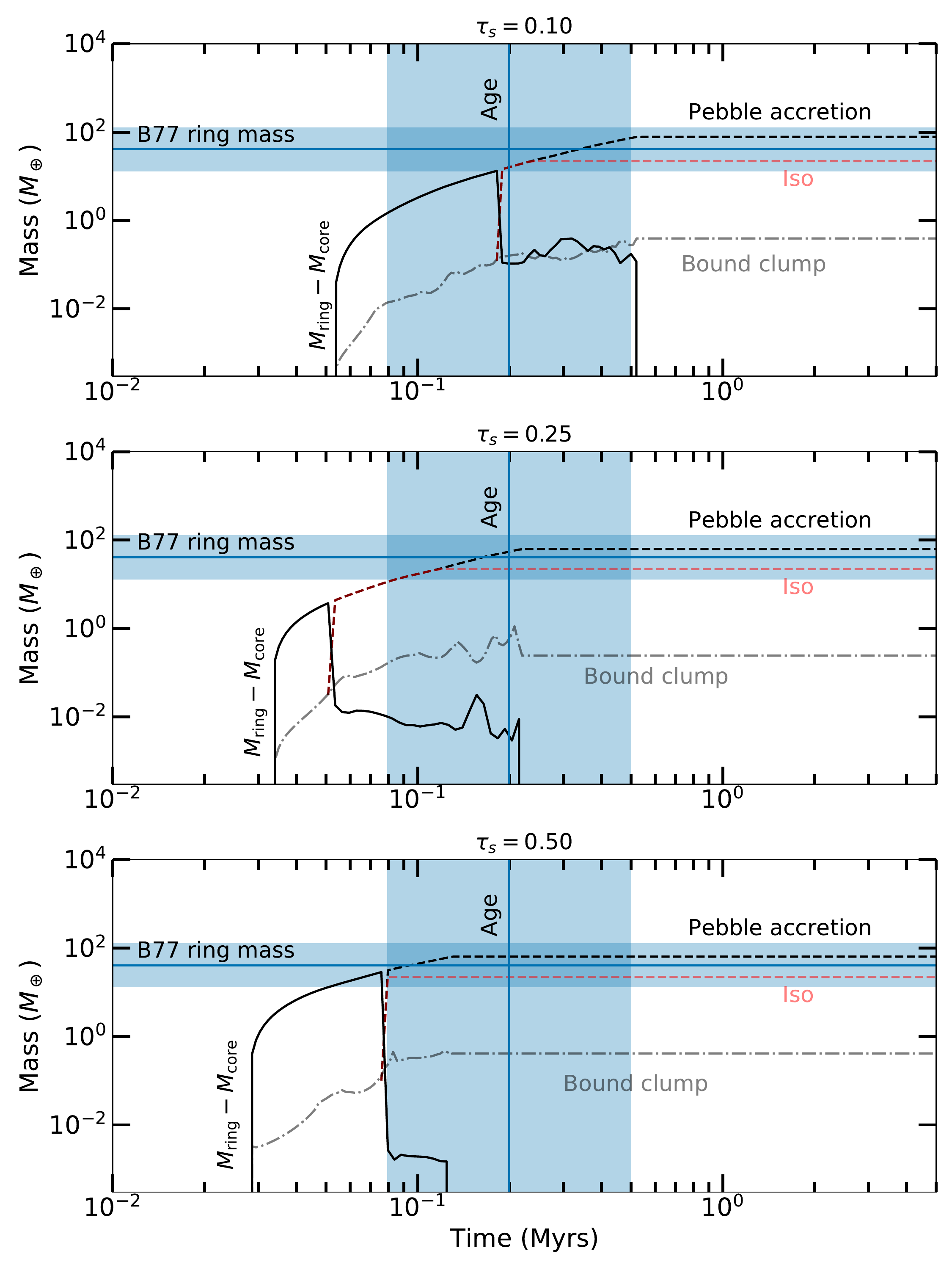}
    \caption{Mass growth of a bound clump under pebble accretion within each dust ring (black dashed line annotated with `Pebble accretion'). Accretion begins when the bound clump (grey dot-dashed line) reaches $M_{\rm settl}$ and the conditions for shear-stability are met (equations \ref{eq:settl_cond1} and \ref{eq:settl_cond2}). For our simulation parameters, the pebble isolation mass (red dashed line; annotated here as `Iso') is $\sim$22.2$M_\oplus$ as computed using the scaling relationship of \citet{Bitsch18}. The black solid line tracks the mass evolution of the ring under the growth by the radial drift of solids exterior to the ring's orbit and the loss of mass to the accreting clump. The vertical and horizontal lines delineate the inferred ages and ring masses of B77 in Elias 24 with the blue bars illustrating 1-$\sigma$ uncertainty \citep{Andrews18,2018ApJ...869L..46D}.}
    \label{fig:Mgrowth_pbump}
\end{figure}

We find that as soon as pebble accretion begins, the clumps can immediately accrete the entire mass of the ring; see the growth tracks illustrated in Figure \ref{fig:Mgrowth_pbump}. To understand these short accretion times, we provide analytic estimates of the timescales to ingest the entire content of the ring ($M_{\rm ring}/\dot{M}_{\rm core}$).

\begin{figure}
    \centering
    \includegraphics[width=0.5\textwidth]{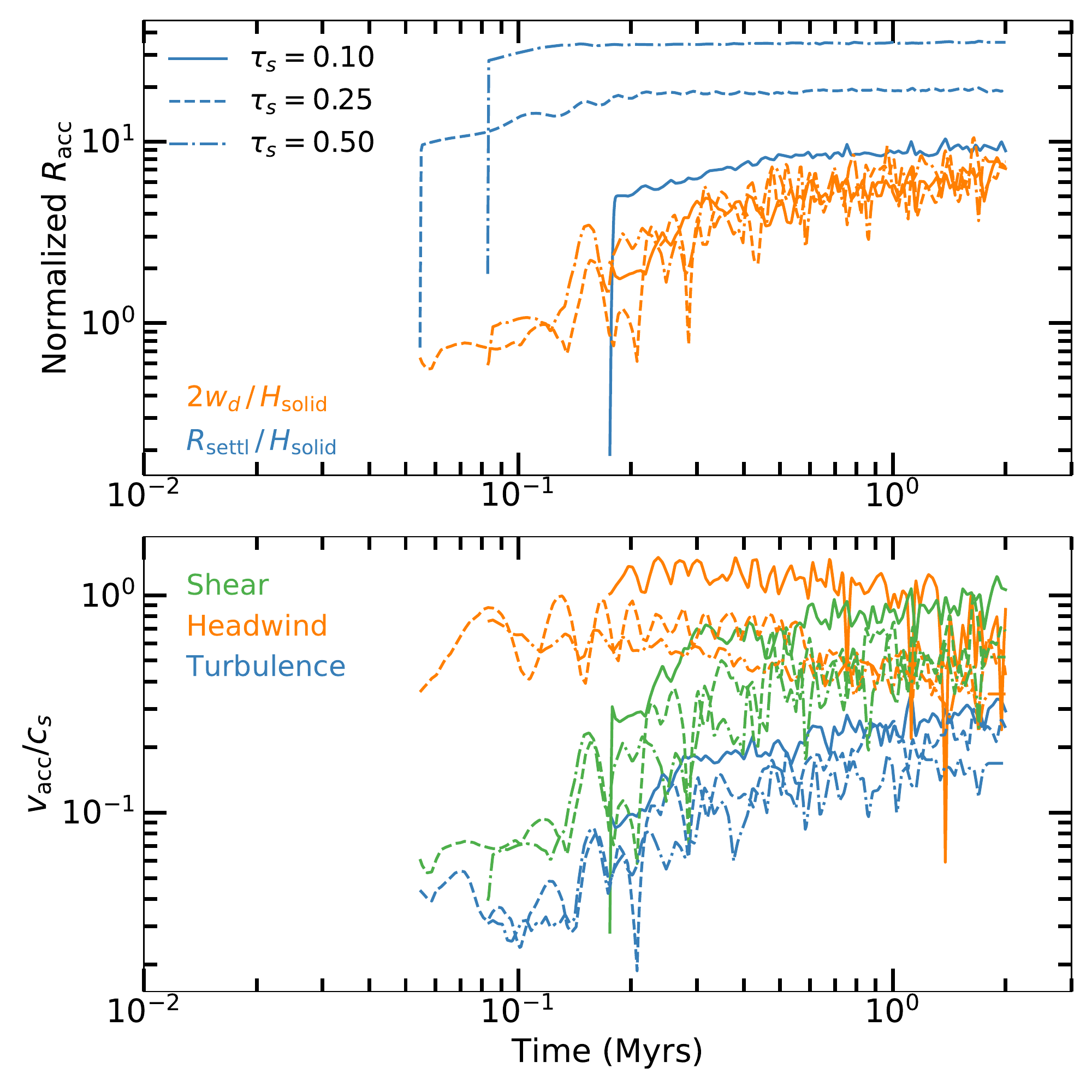}
    \caption{Top: accretion radius under the settling condition ($R_{\rm settl}$, evaluated by solving equation \ref{eq:bacc}) and the width of the dust ring $2w_d$ with respect to $H_{\rm solid}$. Bottom: the relative contribution of shear, headwind (see the text below equation \ref{eq:vacc_comp}), and turbulent random velocities ($v_{\rm rms}$ computed within the ring as defined in equation \ref{eq:v_rms_ring}) to the accretion velocity $v_{\rm acc}$, normalized by the sound speed. In the calculation of the shear velocity, the accretion radius $R_{\rm acc}$ is set to the minimum between $R_{\rm settl}$ and $2w_d$.}
    \label{fig:planet_accr_regime}
\end{figure}

\begin{figure}
    \centering
    \includegraphics[width=0.5\textwidth]{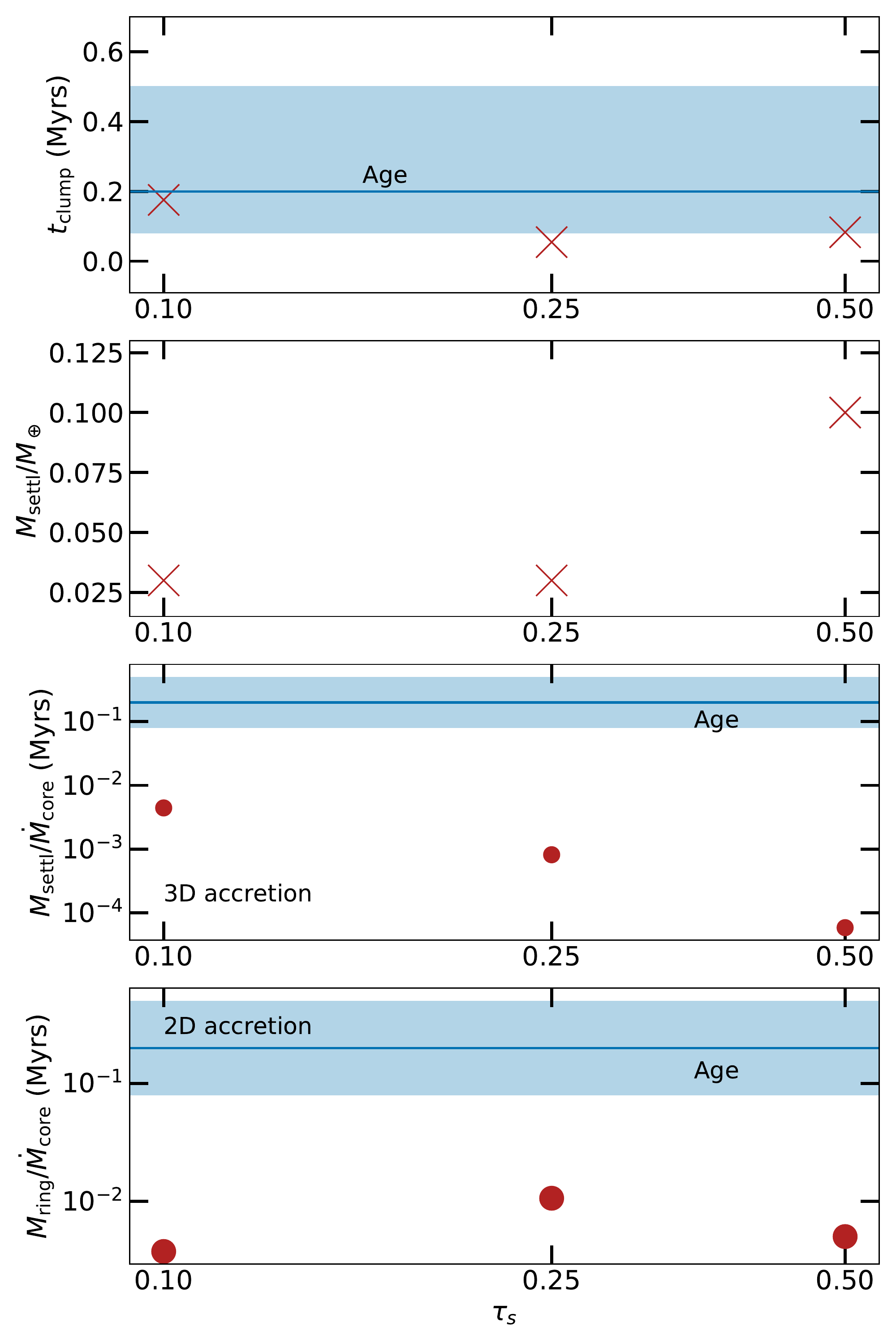}
    \caption{Top: the time at which pebble accretion begins. Second from the top: the initial clump mass for pebble accretion (i.e., minimum mass required for settling accretion). Third from the top: core mass doubling time in 3D accretion ($R_{\rm acc} < H_{\rm solid}$; see equation \ref{eq:Mdot_3d}). Bottom: time to ingest the entire dust ring in 2D accretion ($R_{\rm acc} = 2w_d > H_{\rm solid}$; see equation \ref{eq:Mdot_2d_hw_wd}). For all panels, the horizontal line and the blue bar illustrates the age of Elias 24 and its 1-$\sigma$ uncertainty, respectively.}
    \label{fig:accrtime}
\end{figure}

From Figure \ref{fig:planet_accr_regime}, we infer that the accretion is initially in the 3-dimensional regime ($R_{\rm acc} = {\rm min}(R_{\rm settl}, 2w_d) < H_{\rm solid}$). In this case, we obtain
\begin{equation}
    \dot{M}_{\rm core} = \frac{8\Sigma_{\rm solid}GM_{\rm core}\tau_s}{c_s}\left(\frac{\alpha+\tau_s}{\alpha}\right)^{1/2}
    \label{eq:Mdot_3d}
\end{equation}
by combining equations \ref{eq:Mgrowth_peb} and \ref{eq:vacc_settl}. The core mass grows exponentially in time with the mass doubling time ($M_{\rm core}/\dot{M}_{\rm core}$) being independent of the core mass. As shown in the third panel from the top of Figure \ref{fig:accrtime}, our mass doubling timescales are extremely short as compared to the age of Elias 24, which explains the rapid climb in $M_{\rm core}$ (annotated as `Pebble accretion') seen in Figure \ref{fig:Mgrowth_pbump}.

The transition to the 2-dimensional regime ($R_{\rm acc} > H_{\rm solid}$) is almost immediate in all the runs except for the planet-driven ring of $\tau_s = 0.25$ whose accretion stays in the 3D regime for at least $\sim$0.1-0.2 Myrs. In the 2D accretion, the growth rate depends on the exact behavior of $R_{\rm acc}$ and $v_{\rm acc}$. From Figure \ref{fig:planet_accr_regime}, we infer that the accretion radius of a clump in planet-driven rings will be limited by the width of the dust ring once the accretion enters the 2D regime ($R_{\rm acc} = 2 w_d$) and that the accretion velocity is dominated by the local headwind at all times (while the ring still exists). The accretion rate is then
\begin{equation}
    \dot{M}_{\rm core} = 4M_{\rm ring}\frac{\Omega}{2\pi} \left(\frac{v_{\rm hw}}{a\Omega}\right).
    \label{eq:Mdot_2d_hw_wd}
\end{equation}
From Figure \ref{fig:planet_accr_regime}, we infer $v_{\rm hw} \sim 0.4$--1$c_s$ and since $H/a = 0.05$, $v_{\rm hw}/a\Omega \sim 0.02$--0.05. Since the orbital time at 76.7 AU around 0.78 $M_\sun$ star is $\sim$760 years, equation \ref{eq:Mdot_2d_hw_wd} implies the core is able to accrete the entire ring mass over just 760 years / 4 / 0.05 $\sim 4\times10^3$--$10^4$ years, as shown in the bottom-most panel of Figure \ref{fig:accrtime}. 

All the rings we simulate are capable of creating a stable clump massive enough for pebble accretion, and the mass growth of such a clump is rapid, so much so that we expect the entire ring to be engulfed by the core within the age of the system $\lesssim$1 Myr.
Our result differs from that of \citet{Morbidelli20} who report that planets can only grow at best (defined by them as when the pebble-accreting planetary object is at the radial center of the dust ring) up to $\sim$1$M_\oplus$ in rings such as B77 in Elias 24. One minor difference is our higher $\Sigma_{\rm solid}$, stemming from our tight ring width $w_d$, which accelerates the initial mass doubling in the 3D regime. More crucially, we adopt a larger $\tau_s$ (mainly for the cost of numerical simulation): our smallest $\tau_s=0.05$ as compared to \citet{Morbidelli20} who used $\tau_s = 10^{-3}$ and pebble accretion is expected to be slow for smaller $\tau_s$. 

In 3D pebble accretion, $\dot{M}_{\rm core} \propto \tau_s^{3/2}$ for $\tau_s \ll \alpha$ and we see the expected overall increase in the mass doubling timescale with smaller $\tau_s$ in this regime in Figure \ref{fig:accrtime}.
Once the accretion enters the 2D phase, the trend with $\tau_s$ is less obvious. In this regime ($R_{\rm acc} = 2w_d$, headwind-dominated), $M_{\rm ring}/\dot{M}_{\rm core}$ is independent of $\tau_s$. The core mass doubling time ($M_{\rm core}/\dot{M}_{\rm core}$) however would be $\tau_s$-dependent since at a given time, $M_{\rm ring}$ would be smaller at lower $\tau_s$ due to slower radial drift.
As far as we can tell, the accretion regime used by \citet{Morbidelli20} is 2D, local-headwind dominated regime under the assumption of absolute maximum possible accretion (i.e., the entire dust front undergoing radial drift has the potential to be accreted onto the core, not just the ones that enter $R_{\rm acc}$).

We conclude that the dust rings of characteristics similar to that of DSHARP dataset are, under some circumstances, capable of creating planetary mass objects and if so, these objects would engulf the entire dust within the ring almost instantly. We note that before the ingestion of the entire dust ring, it is likely that these clumps would reach the isolation mass and perturb the ring, perhaps creating another dust ring external to its orbit. To estimate the expected pebble isolation mass, we adopt the scaling relationship of \citet{Bitsch18} assuming $\alpha=10^{-3}$, and the disk aspect ratio to be spatially constant at 0.05 (i.e., equal to our $\Pi$). To calculate the local $dlnP/dlnR$, we take the logarithmic derivative of equation \ref{eq:Pbump} with the background gas surface density to follow $\propto a^{-0.5}$, $\delta_b, w_b = (1.44, 0.96)$ as measured for our planet, and the solid accreting clump to be located at 0.5 scale heights interior to the formal center of the pressure bump as gleaned from Figure \ref{fig:dust_dritribution_2D_planet}. Our expected isolation mass is $\sim$22.2$M_\oplus$, smaller than the median measured mass of the B77 ring but within its 1-$\sigma$ uncertainty. We expect the cores of pebble isolation mass would perturb the surrounding gas (and therefore the dust ring in which the core resides) potentially creating a secondary ring in the outer orbit. Without a significant change in $\tau_{s}$ however, this secondary ring would also be susceptible to near-immediate collapse into a planetary mass object.

The fact that we see these rings over $\sim$1 Myr suggests that such rapid planet formation likely does not happen within the rings. It may be that the dust particles that make up the rings have particularly small $\tau_s$ whose relevant dynamical timescales (e.g., the drift time to fill up the ring and the time of clump formation) are longer (smaller particles have been shown to be preferred solutions to explain low spectral indices \citep[e.g.,][]{Liu19} or the chemical abundances of sulfur-bearing species \citep[e.g.,][]{Harada17}).
We find however that when $\tau_s = 0.05$, planet-driven dust rings tend to be transient as particles are coupled to the advective flow of gas onto the planet, although such transiency may be a feature of our local shearing box approximation.
Alternatively, the measured dust rings may be driven by non-planetary mechanisms that can establish a pressure bump and scatter particles to sufficiently high $v_{\rm rms}$ to reproduce the correct ring width.

As we mentioned previously, the nucleation of dust clumps stable to tidal shear is easier at shorter orbital distances. If typical protoplanetary disks are constantly creating dust rings over a wide range of stellocentric distances and quickly coagulate into planetary objects that create secondary rings, we would expect to see older systems to harbor rings at wider orbits. Such trend however is likely complicated by the intrinsic variance in the size of the protoplanetary disks; in fact, we see no obvious sign of such trend in the DSHARP survey.

We close this section with a comment on the possibility of creating multiple clumps in a single ring. In 3D accretion, all initial clumps would be subject to the same mass doubling time (i.e., $M_{\rm core}/\dot{M}_{\rm core} \propto M_{\rm core}^0$) and so the distribution of relative masses would stay the same. In 2D accretion, from equation \ref{eq:Mdot_2d_hw_wd}, we infer that the mass doubling timescale of a core would lengthen for massive cores. If multiple clumps form in a given ring, the final masses would then approach similar values \citep{Kretke14}. As multiple planetary objects would be placed within a narrow range of orbital distances, their orbits would likely become unstable causing either mergers or ejecta (most likely ejecta at the large orbital distances of DSHARP rings).

\section{Summary and conclusions} \label{sec:conclusions}

Using 2-dimensional (radial-azimuthal plane) shearing box simulations, we studied the interaction between an inward flux of dust particles and gas in a pressure bump established by planet-driven perturbations. Unlike previous studies, we constantly supplied dust particles from the right edge of the simulation box to mimic the inward drift rather than starting with a uniform distribution of particles across the whole box. The main findings are the following:

\begin{enumerate}

    \item Dust particles collect slightly interior to the center of the pressure bump (see Figure \ref{fig:dust_dritribution_2D_planet}). Within the trap, dust particles distribute initially in non-axisymmetric structures and overtime, transform into more axisymmetric rings. Larger $\tau_s$ particles collect more readily into thinner rings.

    \item Vortices triggered by planet-disk interaction help to collect particles, maintaining $\epsilon_{\rm trap} \sim 0.6$--0.8 at all times for $\tau_s = 0.25$ and 0.5. For smaller $\tau_s$, particles are advected out of the dust ring following the gas flows that are attracted to the planet, reducing significantly the efficiency of the trap (down to $40\%$ for particles of $\tau_s=0.1$, and to $\approx 0\%$ for $\tau_s = 0.05$, see Fig. \ref{fig:eff_planet_bump}).
    
    \item With the high $\epsilon_{\rm trap}$, our dust rings are able to collect enough mass within $\lesssim$1 Myr to explain the inferred masses of typical rings analyzed in the DSHARP survey \citep{Andrews18,2018ApJ...869L..46D}. See Figure \ref{fig:masses_rings}.
    
    \item Dust rings start narrow and widen with time, in accordance with drift-diffusion steady state as grains are excited to larger velocity dispersion. The measured width of dust rings in DSHARP data are similar to our simulated rings (see Figure \ref{fig:dust_wd_time}) at small $\tau_{s} = 0.1$.
    
    \item At their maximal size set by the particle disk scale height (assuming $\alpha=10^{-3}$), all our simulated rings are expected to nucleate dust clumps that are gravitationally bound against turbulent diffusion but for the smallest $\tau_{s}$, their clumps are in danger of being sheared apart (see Figure \ref{fig:mass_clumps}). Smaller planetesimals (e.g., smaller than Ceres) may still form.
    
    \item Dust rings made of large particles ($\tau_{s} \geq 0.1$) can nucleate bound and stable clumps massive enough to trigger pebble accretion and such clumps are expected to undergo rapid mass growth ingesting the entire dust content within the ring over timescales $\lesssim$1 Myr. 

\end{enumerate}

The fact that we see concentric dust rings in many of protoplanetary disks imaged with ALMA suggests that the formation of planetary bodies in these rings must be either a rare or a slow process, at least at the wide orbits that are accessible to current interferometric imaging technology. The expected rarity of such wide-orbit planets is in agreement with the statistical analyses from direct imaging \citep[e.g.,][]{Nielsen19} and long baseline radial velocity surveys \citep[e.g.,][]{Fulton21} that suggest gas giant occurrence rate is peaked at $\sim$1--10 AU beyond which it drops.\footnote{We cannot yet rule out the possibility that smaller planets may exist in more abundance at these large orbits \citep[see the hints from microlensing surveys e.g.,][with the caveat that such surveys are more sensitive to M dwarf host stars and orbital separations $\lesssim$10 AU]{Suzuki18} but keeping these planets small and sub-Jovian would be much more natural if they have assembled late rather than early.} 

From our findings, we infer that the real-life disk rings are likely composed of particles of small $\tau_s < 0.1$ so as to delay the creation of dense dust rings, the nucleation of massive planetesimals, and therefore the onset of core growth. One issue with such a solution is that these small particles are not expected to remain in dust rings for long when they are perturbed by a planet. The transient nature of $\tau_s \lesssim 0.05$ rings we found with GIZMO needs to be verified with global disk simulations. We also ignored the planet's gravity on dust particles in order to isolate the dust-gas dynamics. However, as verified in a subset of cases we simulated, turning on planet's gravity acting on dust could cause a stronger leak of particles, reducing the efficiency of the trap and rendering dust rings as transient substructures, particularly for particles of $\tau_s \leq 0.1$, which are already affected by gas inflows into the planet due to the strong coupling with the gas. 

Given the difficulty in maintaining the dust ring against advection at low $\tau_s$ and against engulfment by a planetary object embedded within the ring at high $\tau_s$, the origin of dust rings we see in protoplanetary disks may trace to non-planetary mechanisms. If these rings are the sites of planet formation, then we expect the inner rings to rapidly collapse into a planet or planets first, potentially creating another ring outside their orbits. Under this hypothesis, dust rings would appear at systematically wider orbits for older systems. A larger sample than what we currently have that spans a wider range of ages to search for a trend between the ring location and age may help distinguish between the different origin channels of dust rings.

\begin{acknowledgments}
We thank the anonymous referee for providing a careful report that helped to improve the manuscript. We also thank Ruobing Dong, Jonathan Squires, and Andrew Youdin for helpful discussions and Ge (Wendy) Chen for providing preliminary analyses. E.J.L. gratefully acknowledges support by the Sherman Fairchild Fellowship at Caltech, by NSERC, by le Fonds de recherche du Québec – Nature et technologies (FRQNT), by McGill Space Institute, and by the William Dawson Scholarship from McGill University. J.R.F. acknowledges support by a Mitacs Research Training Award, a McGill Space Institute (MSI) Fellowship, and thanks the Department of Applied Mathematics at the University of Colorado Boulder, for hospitality. 
Support for PFH was provided by NSF Research Grants 1911233, 20009234, 2108318, NSF CAREER grant 1455342, NASA grants 80NSSC18K0562, HST-AR-15800. This research was enabled in part by support provided by Calcul Qu\'ebec (calculquebec.ca) and Compute Canada (www.computecanada.ca). 
\end{acknowledgments}

\vspace{5mm}

\bibliography{references}{}
\bibliographystyle{aasjournal}

\end{document}